\documentclass[journal,draftcls,onecolumn,12pt,twoside]{IEEEtran}
\renewcommand{\baselinestretch}{1.58}
\usepackage{subfigure}
\usepackage{moreverb}
\usepackage{epsfig}
\usepackage{amsmath,amssymb,amsthm,mathrsfs,amsfonts,dsfont}
\usepackage{adjustbox,lipsum}
\usepackage{algorithm,algorithmic}
\usepackage{amsfonts}
\usepackage{epsfig}
\usepackage{amssymb}
\usepackage{cleveref}
\usepackage{amsmath}
\usepackage{amsthm}
\usepackage{subfigure}
\usepackage{multirow}
\usepackage{rotating}
\usepackage{graphicx}
\usepackage{tabularx}
\usepackage{array}
\usepackage{anyfontsize}
\usepackage{color,soul}
\usepackage{graphicx,dblfloatfix}
\usepackage{epstopdf}
\usepackage{blindtext}
\usepackage{amsmath}
\usepackage{amsthm,amssymb,amsmath,bm}
\usepackage{subfigure}
\usepackage{amsfonts}
\usepackage{epsfig}
\usepackage{amssymb}
\usepackage{amsmath}
\usepackage{cite}
\usepackage{graphicx}
\usepackage{fancyhdr}
\usepackage{subfigure}
\usepackage[subfigure]{tocloft}
\usepackage[font={small}]{caption}
\usepackage{subfigure}
\usepackage{tabularx}
\usepackage{cite}

\begin{document}

\title{ Joint Dynamic
	Pricing and Radio Resource Allocation Framework for IoT Services}
\author{Mohammad Moltafet, Atefeh Rezaei,
        Nader Mokari,
         Mohammad~Reza~Javan,
         Hamid Saeedi, and Hossein Pishro-Nik
\thanks{Mohammad Moltafet is with Center for Wireless Communication (CWC), University of Oulu, Oulu, Finland. Atefeh Rezaei, Nader~Mokari, and Hamid Saeedi are with the Department of Electrical and Computer Engineering, Tarbiat Modares University, Tehran, Iran. Mohammad~R.~Javan is with the Department of Electrical and Robotics Engineering, Shahrood University, Shahrood, Hossein Pishro-Nik is with University of Massachusetts, Amherst, MA, USA.}
}

\maketitle
\vspace{-2cm}
\begin{abstract}
In this paper, we study the problem of resource allocation as well as pricing in the context of Internet of things (IoT) networks. We provide a novel pricing model for IoT services where all the parties involved in the communication scenario as well as their revenue and cost are determined. We formulate the resource allocation in the considered model as a multi-objective optimization problem where in addition to the resource allocation variables, the price values are also optimization variables. To solve the proposed multi-objective optimization problem, we use the scalarization method which gives different Pareto optimal solutions.  We solve the resulting problems using the alternating approach based on the successive convex approximation (SCA) method which converges to a local solution with few iterations. We also consider a conventional approach where each entity tries to maximize its own revenue independently. Simulation results indicate that by applying the proposed joint framework, we can increase the total revenue compared to the conventional case while providing an almost complete fairness among the players. This is while the conventional approach fails to provide such a fairness.
\end{abstract}

\section{Introduction}

\normalsize

\subsection{Motivation}

Internet of Things (IoT) is a framework that allows billions of smart devices to be connected to the Internet \cite{7544470}. Such
devices are able to operate and transmit data to other systems with minimal or without any human interaction. The development of IoT has greatly influenced many areas, and many IoT applications have been implemented to improve quality of life in different aspects such as health care, transportation, and manufacturing \cite{8058399}.
There are different 
business models of wireless network virtualization which are described in \cite{6887287} as two-level and four-level models. In two-level model,
mobile network operators (MNO) and service provider (SP) act as logical players after wireless network
virtualization. All of the infrastructures and physical resources are operated by MNOs based on virtualization decisions. SPs operate on virtual
resources to offer end-to-end services to end users. In four-level model, the roles of MNO and SP are decoupled into more specialized tasks, i.e., MNO consists of InP and mobile
virtual network provider (MVNP), and SP consists of mobile virtual network
operator (MVNO) and SP where MVNO assigns the virtual resources to
SPs and SP concentrates on providing services to its end users based on MVNOs decisions.
In this paper, the MNO
becomes InP and SPs will create and deploy
the virtual resource to provide end-to-end
services based on two-level model. 
The core of most IoT systems contains smart wireless sensors that can collect data from the environment and convey such data to the central controllers, referred to as IoT service provider (ISP), for further processing \cite{7544470}. The entity that owns such sensors is referred to as the sensor device owner (SDO).
In addition to SDO and ISP, we can generally consider 4 other essential units: infrastructure providers (InPs), the regulatory, power supplier, and end users. InP provides the required infrastructures and equipments for the communications of different ISPs and lends bandwidth to them, regulatory lends bandwidth to different InPs, and
power supplier provides the necessary electrical power in base stations (BS) and sensors. Finally, the ISP processes the raw data transmitted by sensors and sells them to  end users.
As all of these mentioned entities in an IoT network have their own technical and financial interests,  reaching a financial resource sharing agreement among them
is usually a challenging task. Moreover, most small SDO's and ISPs may not have enough knowledge on the technical and financial details of the service provided by the large InPs and may be billed by the InP's at unreasonable rates. The lack of a transparent market and solid pricing model is one  of the main barriers that prevent IoT from becoming pervasive. 
Therefore, a solid trading/pricing model is necessary to regulate such deals among InPs, ISPs, SDO, and end users. 
 Moreover, IoT in fifth generation (5G) network is required to be able  support massive machine type communication. Massive machine type communication requires enormous amounts of connectivity capability and high spectral efficiency.

\subsection{Related Work}
There are a number of works in wireless networking literature that use pricing methods to model the trade-offs among different entities \cite{7437020,7510922}. Examples include secondary and primary operators in cognitive radio networks \cite{7725543}, device to device (D2D) communications \cite{8014457}, and different base stations in heterogeneous networks \cite{6678362}. The existing literature focuses on bandwidth as the resource to be traded \cite{7811056}. 

In D2D communication, user equipments  transmit data signals to each other over a direct link using the cellular resources rather than the BS to improve bandwidth efficiency. In fact, existing cellular users can sustain their network resources by switching to the D2D mode. Therefore, there must be some incentive  or reward for the D2D users to make them interested to do so. This requires proper pricing strategies for operators to obtain maximum possible profit as shown in \cite{Kebreaei2014, li2016truthful, xu2012resource, xu2012interference}.
In a cognitive  radio network, the primary cellular network owns the licensed spectrum while the secondary users  attempt to dynamically utilize the spectrum. In most cases, such dynamic occupation requires users to pay for the services they get from the primary network through direct billing or by serving as primary users' (PU) relays. Thus, the spectrum becomes a special kind of commodity in a CRN \cite{jiang2015network}. Consequently, a large amount of research has been done to provide different pricing strategies to accommodate efficient spectrum sharing  \cite{xing2007price, ileri2005demand, niyato2008, gandhi2007general, niyato2007hierarchical, simeone2008spectrum, niyato2008market, Sartono2009, jiang2014optimal}.
The idea of heterogeneous networks in which low-cost small cells (e.g., microcells, picocells, and femtocells) with small coverage areas and
low transmission powers are deployed is a promising one for improving the efficiency of spectrum utilization in cellular networks. In this regard, pricing schemes have been considered in  \cite{niyato2007wireless, yun2012economic, kang2012price, shetty2009economics, duan2013economics, yi2012spectrum, chen2012utility, zhu2014pricing}. 

Few papers in the literature have considered pricing schemes for resources other than spectrum. In D2D communications, power has been considered as a subject of trading  in \cite{wangmobile}. In \cite{xu2015optimal} a D2D communication framework is considered in which the authors design a power-pricing framework based on the principle of the Stackelberg game. In \cite{yang2011truthful}, relay servers are a subject of pricing where sellers offer cooperative services at the cost of resources such as power by way of auction. 

As far as pricing in IoT is concerned, only few works exist in the literature \cite{niyato2016smart, niyato2016market, niyato2015optimal, al2013priced,  niyato2018iot}. As a simple market, \cite{niyato2016smart} investigates the pricing scheme in a business model of IoT with three participants: multiple sensing data owners, service providers, and users. With similar market components, \cite{niyato2016market} proposes an economic model in big data and IoT in which the authors use the classification-based machine learning algorithms to define the generic utility function of data.
Then, using a Stackelberg game, the optimal raw data selling price is obtained. Service management of an IoT device has been investigated in \cite{niyato2015optimal} where the Markov decision process  is used to model an optimization framework in order to obtain an optimal policy for the device owner. This policy considers energy transfer and bid acceptance, and attempts to maximize the reward, defined as the revenue from a winning bid and the costs paid for energy transfer.  The authors of \cite{niyato2018iot} consider a cloud based system including  IoT subscriber and  propose a threshold based
approach to  decide the   pricing and allocation
of virtual
machines to sequentially arriving requests in order to maximize
the revenue of the cloud service provider over a finite time horizon. 

  The authors of \cite{Kiani2017} propose a hierarchical mobile edge computing architecture based on the  LTE advanced networks. They study two time scale
 		mechanisms to allocate the computing and communications resources. In the computing resources allocation, they consider
 		an auction-based pricing model to maximize the utility of the service provider where the price of each  virtual machine is updated at the beginning of each frame. To solve this problem, they apply a heuristic algorithm.
 		Moreover, they propose a centralized optimal solution based on Lagrange
 		multipliers for the bandwidth allocation. 
 		The authors of \cite{lk7888924}  consider a  fog computing based system as an appropriate choice to provide low latency services. The considered
network consists of a few data service operators 
each of which controls several fog nodes. The fog nodes provide the required
data service to a set of  subscribers. 
They formulate a Stackelberg game to study the
pricing model for the data service operators as well as the resource allocation
problem for the subscribers. They proposed a many-to-many matching game
to investigate the pairing problem between data service operators and
fog nodes. Moreover, they applied another layer of
many-to-many matching between the  paired fog nodes and
serving data
service subscribers.


Despite its necessity, there is no pricing platform in the IoT literature that can comprehensively address the idea of resource sharing/trading  which can happen at multiple levels. 
		 Motivated by the aforementioned discussion, the objective of this paper is to provide an end to end dynamic pricing and power and subcarrier allocation framework in IoT systems that facilitates reaching an agreement between the network elements and in particular, SDO, InPs,  ISPs, and end users, and to create transparency in such agreements.
\subsection{Our Contributions}

To address the conflicting interests of different players, we have to take into account each entity's objective. Consequently, there are multiple objectives in the network design process which should be optimized simultaneously. Since we have both integer and continuous variables, in  order to jointly maximize the revenues of major players in the proposed pricing model, we use the multi-objective approach which  is a powerful tool that can address such a scenario.

The novelty of the proposed model is two-fold: 
\begin{itemize}
\item A novel comprehensive framework including all the major players and different levels of resource sharing is provided. To the best of our knowledge, no prior work exists that addresses this complex and multilevel pricing structure of IoT systems. 

\item  To solve the proposed end to end optimization problem, several tools including multi-objective optimization, convex optimization and relaxation methods are combined in the framework of wireless communications.

\end{itemize}

We evaluate the performance of the proposed model for different values of the network parameters using simulations. Moreover, we compare the performance of the proposed algorithm  to the conventional approach, in which resource allocation and pricing are disjointly performed by each IoT player. 
From simulations, we can find out that the proposed joint approach leads to much more fairness than the conventional one as the values of the utility functions of involving players are much closer to each other than those of conventional approach.

The organization of the paper is as follows: In Section \ref{Section system model}, the system model and the pricing scheme are presented. In Section \ref{Section Problem Formulations}, problem formulation is provided. Solution of the proposed problem is provided in Section \ref{Section Solution} and simulation results are in Section \ref{section simulation results}. Finally, the paper is concluded in Section \ref{conclusions}.

%

\section{System Model}\label{Section system model}

\subsection{Network Model}\label{SubSection network model}
\begin{figure*}
 	\centering
 	\includegraphics[width=.73\textwidth]{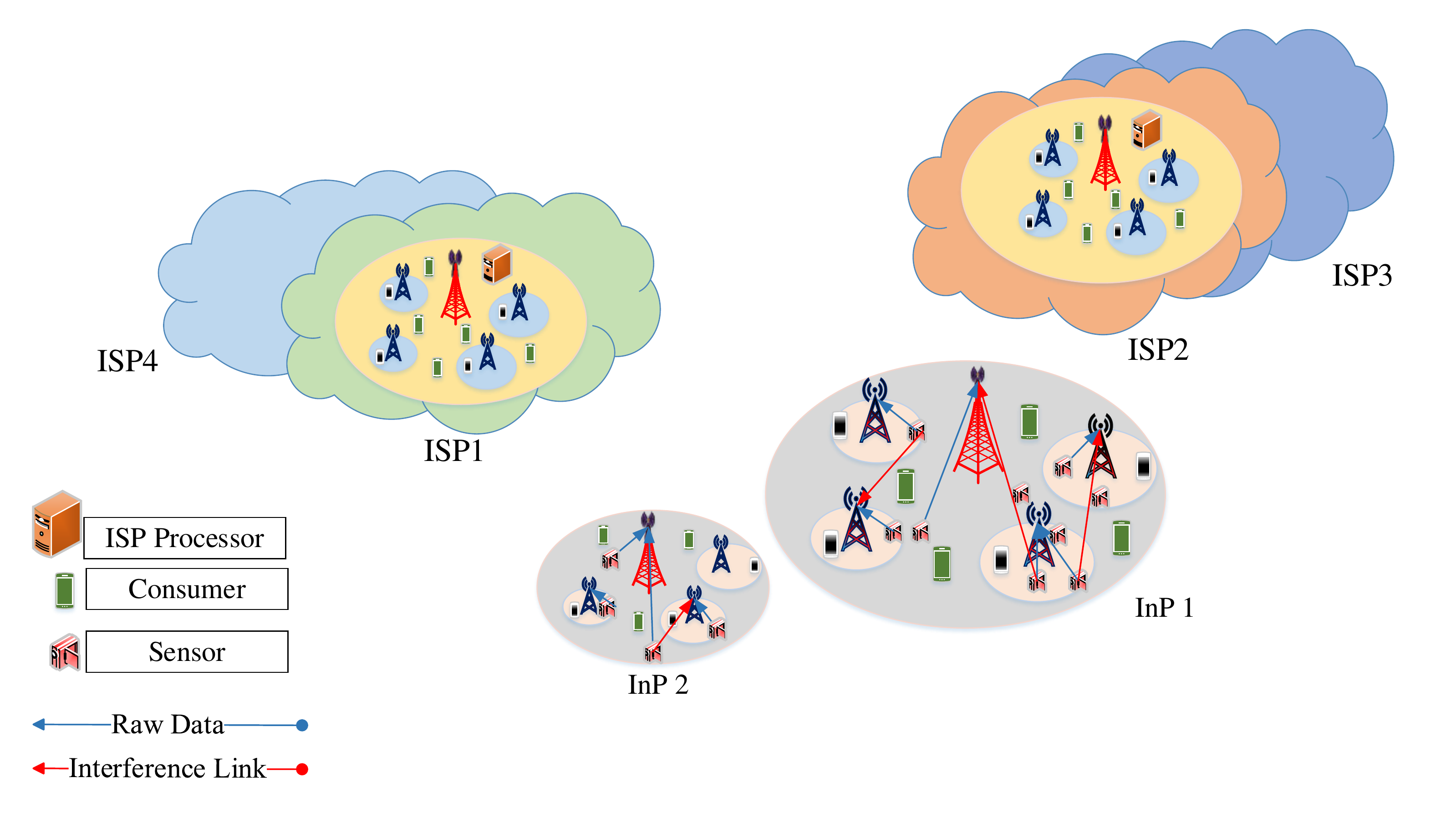}
 	\caption{Uplink transmission from sensors to BSs.}
 	\label{Fig uplink trans}
 \end{figure*}

\begin{figure*}
	\centering
	\includegraphics[width=.73\textwidth]{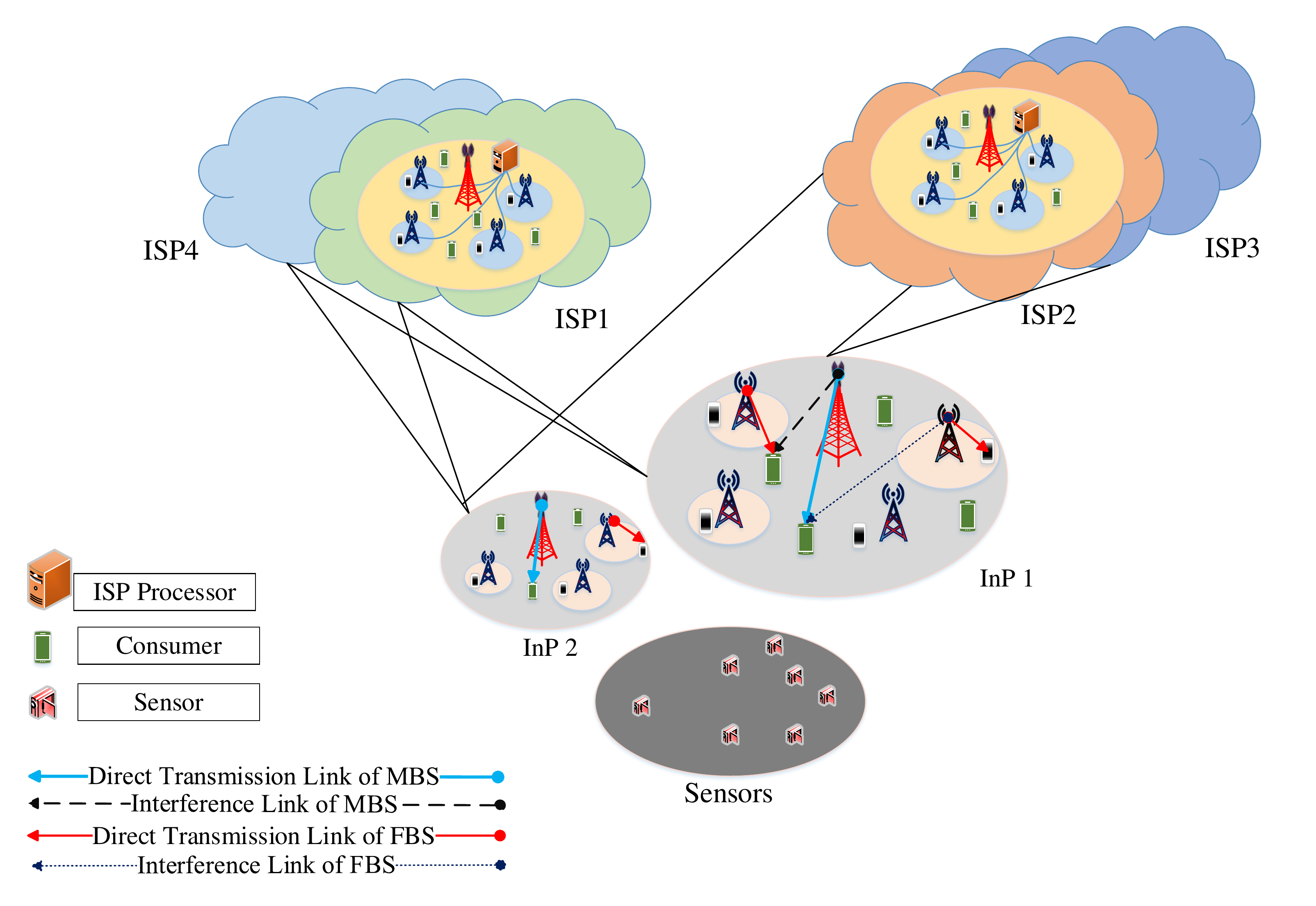}
	\caption{Downlink transmission from BSs to users.}
	\label{Fig downlink trans}
\end{figure*}

\allowdisplaybreaks
Consider a scenario with $U$ users, each of which belongs to one ISP and acts as an IoT service consumer, and  $I$ InPs, where InP $i$ has $B_i$ BSs. On the other hand, there are $S_{b_i}$ sensors in the coverage area of BS $b_i$ who sell the raw data to ISPs. Also each InP framework contains one macro base station (MBS) and few femto base stations (FBSs). In this paper, we focus on both the uplink  and downlink transmission of raw data which is based on frequency division duplex (FDD). The uplink transmission of raw data is the transmission  from sensors to BSs which is shown at Fig. \ref{Fig uplink trans}. The downlink transmission is the transmission of processed data from BSs to users which is shown at Fig. \ref{Fig downlink trans}. The data processing is performed at the private cloud of each ISP\footnote{Note that in our model, the whole infrastructure, provided by multiple InPs, is virtually divided into several virtual networks
	over each of which an ISP provides its services. In this context, the ISPs can be considered as MVNOs.} which is connected to all BSs.  
We denote the set of InPs by $\mathcal{I}=\{1,2,\dots,I\}$, set of BSs in InP $i$ by $\mathcal{B}_i=\{1,2,\dots,B_i\}$, and set of sensors by $\mathcal{S}= \bigcup\limits_{b_i \in \mathcal{B}_i} \mathcal{S}_{b_i}$, where $\mathcal{S}_{b_i}$ is the set of sensors in cell $b_i$ whose cardinality is $S_{b_i}$. Besides, there are $U$ users with the set of $\mathcal{U}=\{1,2,\dots,U\}$, each of which can be associated to only one BS in the network. Moreover, each user is served by one ISP.
The set of ISPs is denoted by $\mathcal{V}=\{1,2,\dots,V\}$. Furthermore, the set of users served by ISP $v$ is denoted by $\mathcal{K}_v$. Hence, we have $\mathcal{U}=\bigcup\limits_{v \in \mathcal{V}} \mathcal{K}_v$.
 The frequency bandwidth of downlink and uplink wireless channels in each InP $i$ are denoted by $W^\text{Dn}_i$ and $W^\text{Up}_i$, respectively.
We assume that different InPs use non-overlapping bandwidth. Within each InP$_i$, the bandwidth is divided between downlink and uplink transmission, denoted by $W^\text{Dn}_i$ and $W^\text{Up}_i$, respectively. These available spectrum  are divided into $N_i$ and $M_i$ subcarriers, respectively. It is assumed that each subcarrier has bandwidth of $W_\text{S}$. The set of downlink and uplink subcarriers in InP $i$ are indicated by $\mathcal{N}_i=\{1,2,\dots,N_i\}$ and $\mathcal{M}_i=\{1,2,\dots,M_i\}$, respectively. 
We utilize the SCMA technique which is one of the main candidates for 5G multiple access techniques \cite{6666156, 7037563,8061028}. We assume that the set of downlink and uplink   codebooks are shown by $\mathcal{DC}_i=\{1,\dots,C_i\}$ and $\mathcal{UC}_i=\{1,\dots,C'_i\}$, respectively, where $C_i$ and $C'_i$ are the number of  downlink and uplink codebooks, respectively. Moreover, the
mapping between downlink subcarriers and codebooks is shown by $q^{c_i}_{n}$, where  $q^{c_i}_{n}=1$ if codebook $c_i$ consists of subcarrier $n$, and otherwise  $q^{c_i}_{n}=0$. In a same way, for uplink case we define $q'^{c'_i}_{m}$ to show the mapping between uplink  subcarriers and codebooks. It should be noted that for both uplink and downlink cases, we assume  that the mapping between codebooks and subcarriers is known and fixed.
Let  $p^{c_i}_{b_i,u}$  denote the downlink transmit power of BS $b_i \in \mathcal{B}_i$ to user $u$ on codebook $c_i \in \mathcal{DC}_i$, and $p'^{c'_i}_{b_i,s_{b_i}}$ denotes the uplink transmit power of sensor $s_{b_i}$ to BS $b_i$ on codebook $c'_i$. The channel  gain between BS $b_i$ and user $u$ on subcarrier $n$, and between BS $b_i$ and sensor $s_{b_i}$ on subcarrier $m$ are determined by $h^{n}_{b_i,u}$ and $h^m_{b_i,s_{b_i}}$, respectively. The codebook assignment indicators are expressed by $\rho^{c_i}_{b_i,u},\rho'^{c'_i}_{b_i,s_{b_i}} \in \{0,1\}$. Note
that  $p^{c_i}_{b_i,u}$ is assigned to   subcarrier $n$  based
on a given proportion $0\le\lambda^{b_i}_{n,c_i}\le 1$, indicated
based on codebook design which satisfies $\sum_{n \in \mathcal{C}_i}\lambda^{b_i}_{n,c_i}=1$ where $\mathcal{C}_i$ is the set of subcarriers in codebook $c_i$. In a same way, $p'^{c'_i}_{b_i,s_{b_i}}$ is assigned to   subcarrier $m$ based on a given proportion  $\lambda'^{b_i}_{m,c_i}$. The definitions of all variables are summarized in Table \ref{table-00}.

\begin{table}[h]
		\caption{NOTATIONS}
	\label{table-00}
	\begin{tabular}{| c| l|l|l|}
		\hline
		Notation& Description&$\boldsymbol{L}^\text{Power,BS}$&$[L^\text{Power,BS}_{b_i}]$\\\hline
	  $\phi^\text{InP,Power}_{i}$&Income of InP $i$ from lending power to ISPs&$\boldsymbol{p}^\text{BS}$ & $[p^{c_i}_{b_i,u}]$ \\\hline
	 $\phi^\text{InP,BW}_i$& 	income of InP $i$ from lending spectrum to ISPs and sensors&$\boldsymbol{p}^\text{Sens}$&  $[p^{c'_i}_{b_i,s_{b_i}}]$\\\hline
		 $\psi^\text{InP,Power}_{i}$& Cost of InP $i$ for buying power from power suppliers&$\boldsymbol{\rho}^\text{BS}$& $[\rho^{c_i}_{b_i,u}]$   \\\hline
		$\psi^\text{InP,BW}_i$ &Cost of buying bandwidth from regulatory with unit price  $C^\text{BW}_i$ &$\boldsymbol{\rho}^\text{Sens}$&  $[\rho'^{c'_i}_{b_i,s_{b_i}}]$\\\hline
		$\psi^\text{ISP,Power}_{v}$ & Cost of buying power from InPs at ISP $v$&$\boldsymbol{L}^\text{BW}$&  $[L^\text{BW}_i]$\\\hline
		$\Phi^\text{InP}_\text{tot}$ & Total revenue of InPs in the network &$\boldsymbol{L}^\text{Sens,Data}$& $[L^\text{Sens,Data}_{v,s_{b_i}}]$\\\hline
		$\psi^\text{ISP,BW}_{v}$ & cost of buying bandwidth from InPs at ISP $v$&$\boldsymbol{L}^\text{Up,Rate}$&  $[L^\text{Up,Rate}_{s_{b_i}}]$\\\hline
	$\psi^\text{ISP,data}_{v} $ &  Price of buying the raw data from sensors at each ISP $v$ &$\boldsymbol{L}^\text{Dn,Rate}$& $[L^\text{Dn,Rate}_v]$\\\hline
		$\phi^\text{ISP,rate}_{v}$ & Income of ISP $v$ originating from the data service &$\boldsymbol{\alpha}$&  $[\alpha_{s_{b_i},u}]$\\\hline
		$\phi^\text{ISP,data}_{v}$ & Income of ISP $v$ from selling the processed data to users &$\boldsymbol{L}^\text{Reserv,User}$&  $[L^\text{Reserv,User}_{v,u}]$\\\hline
		$q^{c_i}_{n}$ &$q^{c_i}_{n}=1$ if codebook $c_i$ consists of subcarrier $n$,   $q^{c_i}_{n}=0$ otherwise&$\mathcal{I}$ & Set of InPs\\\hline
		$q'^{c'_i}_{m}$ &$q'^{c'_i}_{m}=1$ if codebook $c'_i$ consists of subcarrier $m$,   $q^{c_i}_{m}=0$ otherwise&$\mathcal{B}_i$& 	Set of BSs in InP $i$\\\hline
		$p^{c_i}_{b_i,u}$ &Downlink transmit power of BS $b_i \in \mathcal{B}_i$ to user $u$ on codebook $c_i$&$\mathcal{S}$& Set of sensors\\\hline
	    $p'^{c'_i}_{b_i,s_{b_i}}$   &Uplink transmit power of sensor $s_{b_i}$ to BS $b_i$ on codebook $c'_i$&$\mathcal{S}_{b_i}$ &Set of sensors in cell $b_i$\\\hline
		$h^{n}_{b_i,u}$&Channel  gain between BS $b_i$ and user $u$ on subcarrier $n$&$\mathcal{U}$ &Set of  users\\\hline
		$h^m_{b_i,s_{b_i}}$    & Channel  gain between BS $b_i$ and sensor $s_{b_i}$ on subcarrier $m$&$\mathcal{V}$ & Set of ISPs\\\hline
		$\rho^{c_i}_{b_i,u}$    &Downlink codebook assignment&	$\mathcal{N}_i$ & Set of downlink  subcarriers\\\hline
		$\rho'^{c'_i}_{b_i,s_{b_i}}$ & Uplin codebook assignment&$\mathcal{M}_i$ & Set of uplink  subcarriers\\\hline
		$\gamma^{c_i}_{b_i,u}$ &SINR at user $u$ from BS $b_i$ on codebook $c_i$&$\mathcal{DC}_i$ & Set of downlink  codebooks\\\hline
		$r^{c_i}_{b_i,u}$ &Data rate at user $u$ from BS $b_i$ on codebook $c_i$&	$\mathcal{UC}_i$ & Set of uplink  codebooks\\\hline
		$\gamma^{c'_i}_{b_i,s_{b_i}} $ &SINR from sensor $s_{b_i}$ to BS $b_i$ on subcarrier $m$&	$\Phi^\text{InP}_i$ & Revenue of InP $i$\\\hline
		$r^{c'_i}_{b_i,s_{b_i}}$ &Data rate at BS $b_i$ from sensor $s_{b_i}$ on codebook $c'_i$&$\Phi^\text{Sens}_{s_{b_i}}$ & Revenue of each sensor $s_{b_i}$\\\hline
		$P^\text{max}_{b_i}$ &Maximum allowable transmit power of each BS $b_i$&	$\Phi^\text{Sens}_\text{tot} $ & Total revenue of sensors\\\hline
		$P^\text{Bat}_{s_{b_i}}$ & Maximum allowable transmit power of each sensor $s_{b_i}$&$\Phi^\text{ISP}_v$&Revenue of each ISP $v$\\\hline
		$R^\text{min}_{\text{Dn},v}$  &Minimum sum data rate of users owned by ISP $v$&$\Phi^\text{ISP}_\text{tot}$&Total revenue of ISPs\\\hline
		$R^\text{min}_{\text{Up},s_{b_i}}$ & Minimum required data rate at each sensor $s_{b_i}$ for uplink  &$\psi^\text{ISP,Uplink}_{v}$&Paid money from ISP $v$ to SDO \\\hline
		$\phi^\text{User,data}_{u}$&Reward of user $u$ for IoT service&&\\\hline
		$\psi^\text{User,rate}_{u}$&Received data rate by users which is offered by ISP $v$&&\\\hline
	\end{tabular}
\end{table}

The received SINR at user $u$ from BS $b_i$ on codebook $c_i$ is given by
\begin{align}\label{SINR Downlink}
\gamma^{c_i}_{b_i,u} = \dfrac{\rho^{c_i}_{b_i,u}\sum_{n\in \mathcal{N}_i}q^{c_i}_{n}\lambda^{b_i}_{n,c_i}p^{c_i}_{b_i,u}|h^{n}_{b_i,u}|^2}{I^{c_i}_{b_i,u}+(\sigma^{c_i}_{b_i,u})^2},
\end{align}
where $I^{c_i}_{b_i,u}$ is obtained by
\begin{align}
I^{c_i}_{b_i,u}=\sum_{b'_i\in \mathcal{B}_i/\{b_i\}}\sum_{u'\in \mathcal{U}}\sum_{n\in \mathcal{N}_i}\rho^{c_i}_{b'_i,u'}q^{c_i}_{n}\lambda^{b'_i}_{n,c_i}p^{c_i}_{b'_i,u'}|h^{n}_{b'_i,u}|^2.
\end{align}
Accordingly, the data rate at user $u$ from BS $b_i$ on codebook $c_i$ is formulated by
$
r^{c_i}_{b_i,u} = \log_2 (1+\gamma^{c_i}_{b_i,u}).
$
The uplink SINR from sensor $s_{b_i}$ to BS $b_i$ on subcarrier $m$ can be expressed by

\begin{align}\label{SINR uplink}
\gamma^{c'_i}_{b_i,s_{b_i}}  = \dfrac{\rho'^{c'_i}_{b_i,s_{b_i}} \sum_{m\in \mathcal{M}_i}q'^m_{b_i} \lambda'^{b_i}_{n,c'_i}p'^{c'_i}_{b_i,s_{b_i}}|h^{m}_{b_i,s_{b_i}}|^2}
{I'^{c'_i}_{b_i,s_{b_i}}+(\sigma'^{c'_i}_{b_i,s_{b_i}})^2},
\end{align}
where $I'^{c'_i}_{b_i,s_{b_i}}$ is given by
\begin{align}
I'^{c'_i}_{b_i,s_{b_i}}=\sum_{b'_i\in \mathcal{B}_i/\{b_i\}}\sum_{s'_{b'_i}\in \mathcal{S}_{b_i}}\sum_{m\in \mathcal{M}_i}\rho'^{c'_i}_{b'_i,s'_{b'_i}}q^{c'_i}_{m}\lambda'^{b'_i}_{m,c'_i}p'^{c'_i}_{b'_i,s'_{b'_i}}|h^{m}_{b_i,s'_{b'_i}}|^2.
\end{align}

The received data rate at BS $b_i$ from sensor $s_{b_i}$ on codebook $c'_i$ is thus formulated by
$
r^{c'_i}_{b_i,s_{b_i}} = \log_2 (1+\gamma^{c'_i}_{b_i,s_{b_i}}).
$
The following two constraints ensure  that each user selects one BS in the network:
\begin{align}\label{cons userassociation dn1}
\rho^{c_i}_{b_i,u} + \rho^{c''_{i'}}_{b_{i'}',u} \leq 1, \forall i,i' \in \mathcal{I}, i' \neq i, b_i \in \mathcal{B}_i , b_{i'}' \in \mathcal{B}_{i'},  u \in \mathcal{U}, c_i \in \mathcal{DC}_i, c''_{i'} \in \mathcal{DC}_{i'},
\end{align}
\begin{align}\label{cons userassociation dn2}
\rho^{c_i}_{b_i,u} + \rho^{c'''_{i}}_{b_i',u} \leq 1, \forall i \in \mathcal{I}, b_i,b_i' \in \mathcal{B}_i , b_i' \neq b_i, u \in \mathcal{U}, c_i,\ c'''_i \in \mathcal{DC}_i.
\end{align}
In SCMA, each subcarrier can at most be reused $K$ times, therefore, the following constraints are enforced for downlink and uplink, respectively: 
\begin{align}\label{SCMAc1}
\sum_{ b_i \in \mathcal{B}_i}\sum_{ u \in \mathcal{U}}\sum_{ c_i \in \mathcal{DC}_i}\rho^{c_i}_{b_i,u}q^{c_i}_{n}\le K, \forall i\in \mathcal{I}, n\in \mathcal{N}_i,
\end{align}
\begin{align}\label{SCMAc2}
\sum_{ b_i \in \mathcal{B}_i}\sum_{ s_{b_i}\in \mathcal{S}_i}\sum_{ c'_i \in \mathcal{UC}_i}\rho'^{c'_i}_{b_i,s_{b_i}}q'^{c'_i}_{m}\le K, \forall i\in \mathcal{I}, m\in \mathcal{M}_i.
\end{align}

Maximum allowable transmit power of each BS $b_i$ and sensor $s_{b_i}$ are denoted by $P^\text{max}_{b_i}$ and $P^\text{Bat}_{s_{b_i}}$, respectively. Hence, we have
\begin{align}\label{cons max power BS}
\sum_{ u \in \mathcal{U}}\sum_{c_i \in \mathcal{C}_i} \rho^{c_i}_{b_i,u} p^{c_i}_{b_i,u} \leq P^\text{max}_{b_i}, \forall i \in \mathcal{I}, b_i \in \mathcal{B}_i,
\end{align}
\begin{align}\label{cons max power user}
 \sum_{c'_i \in \mathcal{UC}_i} \rho^{c'_i}_{b_i,s_{b_i}} p^{c'_i}_{b_i,s_{b_i}} \leq P^\text{Bat}_{s_{b_i}}, \forall i \in \mathcal{I}, b_i \in \mathcal{B}_i, s_{b_i} \in \mathcal{S}_{b_i}.
\end{align}
According to the different user rate requirements of various ISPs, we have the following minimum required data rate constraints as:
\begin{align}\label{cons min rate user dn}
\sum_{ i \in \mathcal{I}} \sum_{ b_i \in \mathcal{B}_i} \sum_{c_i \in \mathcal{DC}_i}  r^{c_i}_{b_i,u} \geq R^\text{min}_{\text{Dn},v}, \forall v \in \mathcal{V}, u \in \mathcal{K}_v,
\end{align}
where $R^\text{min}_{\text{Dn},v}$ is the minimum sum data rate of users owned by ISP $v$. In the same way, we have a minimum data rate constraint for each sensor as:
\begin{align}\label{cons min rate user up}
\sum_{c_i \in \mathcal{UC}_i}  r^{c'_i}_{b_i,s_{b_i}} \geq R^\text{min}_{\text{Up},s_{b_i}}, \forall b_i \in \mathcal{B}_i, s_{b_i} \in \mathcal{S}_{b_i},
\end{align}
where $R^\text{min}_{\text{Up},s_{b_i}}$ is the minimum required data rate at each sensor $s_{b_i}$ for uplink transmission.

\subsection{Pricing Scheme}\label{SubSection Pricing Scheme}
\begin{figure*}
	\centering
	\includegraphics[width=.7\textwidth]{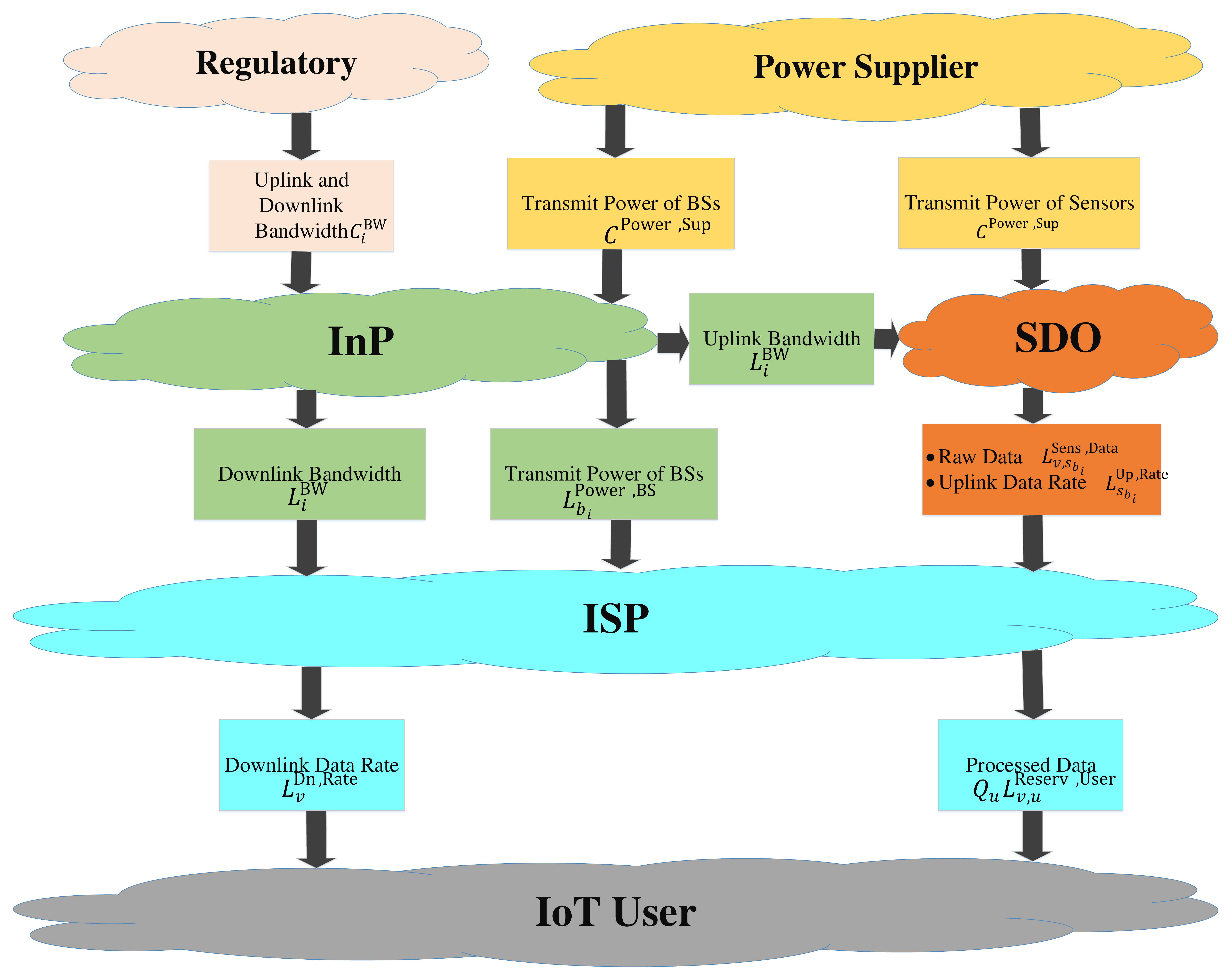}
	\caption{IoT Pricing Scheme.}
	\label{IoT pricing scheme}
\end{figure*}

In the proposed pricing model, each sensor sells its raw sensed data to ISPs. In other words, each ISP collects data from different sensors. Then, ISPs perform a processing based on the determined sets of collected data and sell it to users. Fig. \ref{IoT pricing scheme} illustrates the proposed pricing model of the considered system.
Based on Fig. \ref{IoT pricing scheme}, there are four major players in the proposed IoT pricing scheme: InPs, sensors, ISPs and users. In the following, we present the pricing model of each player.
\subsubsection{InPs}
Each InP leases power and bandwidth from a power supplier and regulatory, respectively, and lends them to ISPs. In addition, each InP $i$ lends its bandwidth to the set of sensors owned by BSs in $\mathcal{B}_i$. Assume that $L^\text{Power,BS}_{b_i}$ let the unit of price of power consumption at BS $b_i$, per Watt.
The income of InP $i$ from lending power to ISPs can be obtained as
\begin{align}\label{lending resources to ISPs1}
\phi^\text{InP,Power}_{i} = \sum_{ b_i \in \mathcal{B}_i} \sum_{ u \in \mathcal{U}} \sum_{c_i \in \mathcal{DC}_i} L^\text{Power,BS}_{b_i} \rho^{c_i}_{b_i,u} p^{c_i}_{b_i,u}.
\end{align}
Let $L^\text{BW}_i$ denote  the unit price for lending downlink and uplink transmission bandwidths, per Hz, to ISPs and sensors. Therefore, the income of InP $i$ from lending spectrum to ISPs and sensors are given by
\begin{align}\label{lending resources to ISPs2}
&\phi^\text{InP,BW}_i = \sum_{ b_i \in \mathcal{B}_i} \sum_{u \in \mathcal{U}} \sum_{c_i \in \mathcal{DC}_i}\sum_{n\in \mathcal{N}_i}q^{c_i}_{n} L^\text{BW}_i \rho^{c_i}_{b_i,u} W_\text{S} + \\& \nonumber\sum_{b_i \in \mathcal{B}_i} \sum_{s_{b_i} \in \mathcal{S}_{b_i}} \sum_{c'_i \in \mathcal{UC}_i} \sum_{m\in \mathcal{M}_i}q'^m_{b_i} L^\text{BW}_i \rho^{c'_i}_{b_i,s_{b_i}} W_\text{S}.
\end{align}
The cost of InP $i$ for buying power from power suppliers with a unit price $C^\text{Power,Sup}$ (per Watt) is obtained by
\begin{align}\label{cost InP power}
\psi^\text{InP,Power}_{i} = \sum_{ b_i \in \mathcal{B}_i} \sum_{ u \in \mathcal{U}}\sum_{c_i \in \mathcal{C}_i} \rho^{c_i}_{b_i,u} p^{c_i}_{b_i,u} C^\text{Power,Sup}, \forall i \in \mathcal{I},
\end{align}
and the cost of buying bandwidth from regulatory with unit price  $C^\text{BW}_i$ (per Hz) can be formulated by
\begin{align}\label{cost InP BW}
\psi^\text{InP,BW}_i = W^\text{Dn}_i C^\text{BW}_i + W^\text{Up}_i C^\text{BW}_i, \forall i \in \mathcal{I}.
\end{align}
The revenue of InP $i$ is thus given by
\begin{align}\label{revenue InP}
\Phi^\text{InP}_i = \phi^\text{InP,Power}_{i} + \phi^\text{InP,BW}_i - \psi^\text{InP,Power}_{i} - \psi^\text{InP,BW}_i.
\end{align}
Hence, the total revenue of InPs in the network is formulated as 
$
\Phi^\text{InP}_\text{tot} = \sum_{i \in \mathcal{I}} \Phi^\text{InP}_i.
$

\subsubsection{Sensors}
Each sensor $s_{b_i}$ has a reservation wage cost $C^\text{Sens,Reserv}_{s_{b_i}}$ to obtain the raw data. Moreover, it leases power from the power supplier and bandwidth from InPs with units price $C^\text{Power,Sup}$ (per Watt) and $L^\text{BW}_i$ (per Hz), respectively, to sell the raw data with the price $L^\text{Sens,Data}_{v,s_{b_i}}$ which is offered by ISP $v$ to sensor $s_{b_i}$ in uplink transmission to ISPs. In doing so, the uplink data rate of each sensor $s_{b_i}$ is the income of the sensor with unit of price $L^\text{Up,Rate}_{s_{b_i}}$ (1/bps).
Accordingly, the revenue of each sensor $s_{b_i}$ can be obtained by
\begin{align}\label{revenue sensor}
&\Phi^\text{Sens}_{s_{b_i}} =
\sum\limits_{v \in \mathcal{V}} \min \{ \sum\limits_{u \in \mathcal{K}_v} \alpha_{s_{b_i},u} , 1 \} L^\text{Sens,Data}_{v,s_{b_i}}
+ \sum\limits_{v \in \mathcal{V}} \sum_{c'_i \in \mathcal{UC}_i} \min \{ \sum\limits_{u \in \mathcal{K}_v} \alpha_{s_{b_i},u} , 1 \} r^{c'_i}_{b_i,s_{b_i}} L^\text{Up,Rate}_{s_{b_i}}
\\& \nonumber-
\min \{ \sum\limits_{u \in \mathcal{U}} \alpha_{s_{b_i},u} , 1 \} C^\text{Sens,Reserv}_{s_{b_i}}-
\sum\limits_{c'_i \in \mathcal{UC}_i} \rho^{c'_i}_{b_i,s_{b_i}} p^{c'_i}_{b_i,s_{b_i}} C^\text{Power,Sup} - \sum_{c'_i \in \mathcal{UC}_i} \sum_{m\in \mathcal{M}_i}q'^m_{b_i}  \rho^{c'_i}_{b_i,s_{b_i}} W_\text{S} L^\text{BW}_i,
\end{align}
where the binary indicator $\alpha_{s_{b_i},u} \in \{0,1\}$ takes value $1$ when the raw data of sensor $s_{b_i}$ is used in the processing data of user $u$ at the cloud of ISP. Note that each ISP buys the data of sensor $s_{b_i}$ at most once.
The total revenue of sensors is thus given by
$
\Phi^\text{Sens}_\text{tot} = \sum_{s_{b_i} \in \mathcal{S}_{b_i}} \Phi^\text{Sens}_{s_{b_i}}.
$

\subsubsection{ISPs}
Each ISP buys power and downlink bandwidth from InP with unit price $L^\text{Power,BS}_{b_i}$ and $L^\text{BW}_i$, respectively. Moreover, it leases the raw data of sensors with price $L^\text{Sens,Data}_{v,s_{b_i}}$ and gives money to sensors for their uplink data rate with the unit price $L^\text{Up,Rate}_{s_{b_i}}$. The income of each ISP is composed of two components. Firstly, they get money for their downlink data rate servicing to users with the unit of price $L^\text{Dn,Rate}_v$ (per bit/s). Secondly, they lend their processed data to users. The price of the processed data which is served to user $u$ from ISP $v$ is $Q_u L^\text{Reserv,User}_{v,u}$, where $Q_u = q \log( 1+\frac{\sum\limits_{i=1}^{I} \sum\limits_{b_i \in \mathcal{B}_i} \sum\limits_{s_{b_i} \in \mathcal{S}_{b_i}} \alpha_{s_{b_i},u}} { \sum\limits_{i=1}^{I} \sum\limits_{b_i \in \mathcal{B}_i} S_{b_i}})$ is the service quality function of a set of sensors whose data is used by user $u$ \cite{7437020}, and $L^\text{Reserv,User}_{v,u}$ is the maximum reservation price that ISP $v$ takes from user $u$. In addition, $q$ is a sensing quality of player that is used to tune service quality received by the users.
In doing so, the income of ISP $v$ originating from the data service given to the its  subscribing users is given by
\begin{align}\label{Reward ISP rate}
\phi^\text{ISP,rate}_{v} =  \sum_{ i \in \mathcal{I}} \sum_{ b_i \in \mathcal{B}_i} \sum_{ u \in \mathcal{K}_v} \sum_{c_i \in \mathcal{DC}_i}\sum_{n\in \mathcal{N}_i}q^{c_i}_{n} L^\text{Dn,Rate}_v W_\text{S} r^{c_i}_{b_i,u}.
\end{align}
Moreover, the income of ISP $v$ from selling the processed data to users in $\mathcal{K}_v$ is obtained by
$
\phi^\text{ISP,data}_{v} = \sum_{ u \in \mathcal{K}_v} Q_u L^\text{Reserv,User}_{v,u}.
$
The cost of buying power from InPs at ISP $v$ is given by
\begin{align}\label{cost ISP power}
\psi^\text{ISP,Power}_{v} = \sum_{ i \in \mathcal{I}} \sum_{ b_i \in \mathcal{B}_i} \sum_{ u \in \mathcal{K}_v} \sum_{c_i \in \mathcal{DC}_i} L^\text{Power,BS}_{b_i} \rho^{c_i}_{b_i,u} p^{c_i}_{b_i,u},
\end{align}
and the cost of buying bandwidth from InPs at ISP $v$ is given by
\begin{align}\label{cost ISP BW}
\psi^\text{ISP,BW}_{v} = \sum_{ i \in \mathcal{I}} \sum_{ b_i \in \mathcal{B}_i} \sum_{u \in \mathcal{K}_v} \sum_{c_i \in \mathcal{DC}_i}\sum_{n\in \mathcal{N}_i}q^{c_i}_{n} L^\text{BW}_i \rho^{c_i}_{b_i,u} W_\text{S} .
\end{align}
The price of buying the raw data from sensors at each ISP $v$ can be formulated by
\begin{align}\label{cost ISP rawdata0}
\psi^\text{ISP,data}_{v}= \sum_{s_{b_i} \in \mathcal{S}_{b_i}} \min \{ \sum\limits_{u \in \mathcal{K}_v} \alpha_{s_{b_i},u} , 1 \} L^\text{Sens,Data}_{v,s_{b_i}}.
\end{align}
Moreover, each ISP $v$ gives money to SDO for their uplink data rates with the total price of
\begin{align}\label{cost ISP rawdata}
&\psi^\text{ISP,Uplink}_{v} = \sum_{i \in \mathcal{I}} \sum_{b_i \in \mathcal{B}_i} \sum_{s_{b_i} \in \mathcal{S}_{b_i}} \min \{ \sum\limits_{u \in \mathcal{K}_v} \alpha_{s_{b_i},u} , 1 \}  r^{c'_i}_{b_i,s_{b_i}} L^\text{Up,Rate}_{s_{b_i}}.
\end{align}
The revenue of each ISP $v$ is thus formulated as follows:
\begin{align}\label{revenue ISP2}
&\Phi^\text{ISP}_v = \phi^\text{ISP,rate}_{v} + \phi^\text{ISP,data}_{v} - \psi^\text{ISP,Power}_{v} - \psi^\text{ISP,BW}_{v} - \psi^\text{ISP,data}_{v} - \psi^\text{ISP,Uplink}_{v}.
\end{align}
Therefore, the total revenue of ISPs is given by
$
\Phi^\text{ISP}_\text{tot} = \sum_{v \in \mathcal{V}} \Phi^\text{ISP}_v.
$

\subsubsection{Users}

Each user is willing to achieve the high quality IoT  and high data rate services with low power and spectrum consumption. Therefore, for each user, rewards and costs are modeled as follows:
\begin{itemize}
  \item Reward 1: IoT service (the processed data) is received by users with unit price $C^\text{Reserv,User}_{u}$. Therefore, reward of user $u$ for IoT service  is given by
      \begin{align}\label{reward user data}
        \phi^\text{User,data}_{u} = Q_u C^\text{Reserv,User}_{u}, \forall u \in \mathcal{K}_v.
      \end{align}
  \item Cost 1: The received data rate by users which is offered by ISP $v$ with unit cost $L^\text{Dn,Rate}_v$ is given  by
                \begin{align}\label{cost user rate}
                 \psi^\text{User,rate}_{u} = \sum_{ i \in \mathcal{I}} \sum_{ b_i \in \mathcal{B}_i} \sum_{c_i \in \mathcal{DC}_i}\sum_{n\in \mathcal{N}_i}q^{c_i}_{n} L^\text{Dn,Rate}_v W_\text{S} r^{c_i}_{b_i,u}, \forall u \in \mathcal{K}_v.
                \end{align}
  \item Cost 2: IoT service price at user $u$ with unit cost $L^\text{Reserv,User}_{v,u}$ which is offered by ISP $v$ to user $u$,  can be obtained by $Q_u L^\text{Reserv,User}_{v,u}, \forall u \in \mathcal{K}_v$.
\end{itemize}
The revenue of each user $u \in \mathcal{K}_v$ is formulated as follows:
\begin{align}\label{revenue User}
\Phi^\text{User}_u = \phi^\text{User,data}_{u} - \psi^\text{User,rate}_{u} - Q_u L^\text{Reserv,User}_{v,u}, \forall u \in \mathcal{K}_v.
\end{align}
Therefore, the total revenue of users is obtained by
$
\Phi^\text{User}_\text{tot} = \sum_{u \in \mathcal{U}} \Phi^\text{User}_u.
$

\section{Proposed pricing Model}\label{Section Problem Formulations}
In this paper, we aim to design a joint uplink/downlink data delivery policy with BS selection at users and determine the efficient value of pricing units offered by each player. Moreover, we obtain the efficient raw data subset selection for processing the collected raw data for each user.
Let us denote $\boldsymbol{p}=[\boldsymbol{p}^\text{BS},\boldsymbol{p}^\text{Sens}]$, $\boldsymbol{p}^\text{BS}=[p^{c_i}_{b_i,u}]$, $\boldsymbol{p}^\text{Sens}=[p^{c'_i}_{b_i,s_{b_i}}]$, $\boldsymbol{\rho}=[\boldsymbol{\rho}^\text{BS},\boldsymbol{\rho}^\text{Sens}]$, $\boldsymbol{\rho}^\text{BS}=[\rho^{c_i}_{b_i,u}]$, $\boldsymbol{\rho}^\text{Sens}=[\rho'^{c'_i}_{b_i,s_{b_i}}]$,
$\boldsymbol{L}^\text{Power,BS}=[L^\text{Power,BS}_{b_i}]$, $\boldsymbol{L}^\text{BW}=[L^\text{BW}_i]$, $\boldsymbol{L}^\text{Sens,Data}=[L^\text{Sens,Data}_{v,s_{b_i}}]$,
$\boldsymbol{L}^\text{Up,Rate}=[L^\text{Up,Rate}_{s_{b_i}}]$, $\boldsymbol{L}^\text{Dn,Rate}=[L^\text{Dn,Rate}_v]$,
$\boldsymbol{\alpha}=[\alpha_{s_{b_i},u}]$, $\boldsymbol{L}^\text{Reserv,User}=[L^\text{Reserv,User}_{v,u}]$,
$\boldsymbol{L}=[\boldsymbol{L}^\text{Power,BS},\boldsymbol{L}^\text{BW},\boldsymbol{L}^\text{Sens,Data},$\\$
\boldsymbol{L}^\text{Up,Rate},  
\boldsymbol{L}^\text{Dn,Rate},\boldsymbol{L}^\text{Reserv,User}]$. 
  Since we have both integer and continuous variables, in  order to jointly maximize the revenues of the different players, we use the multi-objective approach.
  The multi-objective optimization problem of the proposed system model is formulated as 
	\begin{align}\label{MOO010}
	\max_{\boldsymbol{\rho},\boldsymbol{p},\boldsymbol{L},\boldsymbol{\alpha} }
	\{\Phi^\text{ISP}_\text{tot},
	\Phi^\text{InP}_\text{tot}, 
	\Phi^\text{Sens}_\text{tot},
	\Phi^\text{User}_\text{tot}\},
	\text{s.t.}:\hspace{.25cm}
	\eqref{cons userassociation dn1}\text{-}\eqref{cons min rate user up}.
	\end{align}
The proposed optimization problem in \eqref{MOO010} is intractable.
To tackle this issue,  we  adopt the scalarization method. 
By using this method, a multi-objective function can be transformed into  a simple, tractable,  and single objective function \cite{7917236,781910,6924852}. Note that there are many cases in the scalarization methods resulting in different Pareto optimal solutions. The result of solution of  each single objective problem  forms a point of Pareto boundary.   In the following, we exploit two approaches within the scalarization method, namely, max-min and weighted-one algorithms, to solve the  multi-objective optimization problem. The main reasons behind selection of these methods are that the weighted-one is very  simple  and the max-min method achieves the best fairness in the system.  For more clarification please refer to Appendix A.

\subsection{Max-Min Approach}
ISPs, InPs, and SDO are three players which work together to provide IoT  services  for  end users. In this approach, our goal is  to maximize the fairness index among these three players. 
 Consequently,
our max-min optimization problem is formulated as follows: 
	\begin{align}\label{MOO01a}
		\max_{\boldsymbol{\rho},\boldsymbol{p},\boldsymbol{L},\boldsymbol{\alpha} }
		\min_{\text{ISP},\text{InP},\text{Sens},}\{
		 			\Phi^\text{ISP}_\text{tot},
			\Phi^\text{InP}_\text{tot}, 
			\Phi^\text{Sens}_\text{tot}\}+
			\Phi^\text{User}_\text{tot}, 
		\text{s.t.}:\hspace{.25cm}
		\eqref{cons userassociation dn1}\text{-}\eqref{cons min rate user up}. 
	\end{align}
\subsection{Weight-One Approach}
In this approach, our goal is  to maximize the summation of  weighted utilities of ISPs, InPs, sensors, and users. The max-min optimization problem is formulated as follows:

\begin{subequations}\label{problem main}
	\begin{align}\label{obf problem main}
	&\max_{ \boldsymbol{\rho},\boldsymbol{p},\boldsymbol{L},\boldsymbol{\alpha} }
    \sum_{v \in \mathcal{V}} \omega^\text{ISP}_v \Phi^\text{ISP}_v + \sum_{u \in \mathcal{U}} \omega^\text{User}_u \Phi^\text{User}_u + \sum_{i \in \mathcal{I}} \omega^\text{InP}_i \Phi^\text{InP}_i + \sum_{i \in \mathcal{I}} \sum_{b_i \in \mathcal{B}_i} \sum_{s_{b_i} \in \mathcal{S}_{b_i}} \omega^\text{Sens}_{s_{b_i}} \Phi^\text{Sens}_{s_{b_i}}
    \\
    & \text{s.t.}\hspace{.1cm}
	\eqref{cons userassociation dn1}\text{-}\eqref{cons min rate user up}.    \nonumber
	\end{align}
\end{subequations}
where $\omega^\text{ISP}_v$, $\omega^\text{User}_u$, $\omega^\text{InP}_i$ and $\omega^\text{Sens}_{s_{b_i}}$ are the weights tuned based on the priority of $\Phi^\text{ISP}_v$, $\Phi^\text{User}_u$, $\Phi^\text{InP}_i$ and $\Phi^\text{Sens}_{s_{b_i}}$, respectively.
\section{Conventional Approach in IoT Pricing}\label{Conventional Approach in IoT Pricing}
Based on a conventional approach, each player by considering a minimum utility requirement for the other players maximizes its utility. Then, the calculated results of its corresponding price variables 
are reported  to a central unit. Central unit by considering the reported price variables solves a comprehensive joint power and codebook allocation problem for all of the players. Consequently, by the conventional method, price variables are determined by players and power and codebook are assigned centrally.    In the following, the optimization problems which should be solved by different players and the optimization problem which should be solved by the central unit are presented.  It should be noted that each InP can play the role of the central unit to solve the  comprehensive joint power and codebook allocation problem.
\subsection{InP Optimization Problem}
Each InP maximizes its utility function by considering minimum utilities $\Phi^\text{ISP}_0$, $\Phi^\text{User}_0$ and $\Phi^\text{SDO}_0$ for ISPs, users and SDO, respectively. Moreover, for other InPs it considers minimum utility $\Phi^\text{EInP}_0$. The corresponding optimization problem is formulated as:
\begin{subequations}\label{MOO0121}
	\begin{align}\label{MOO01a21}
	&\max_{\boldsymbol{\rho},\boldsymbol{p},\boldsymbol{L},\boldsymbol{\alpha}}
	\Phi^\text{InP}_i,
	\\&\nonumber
	\text{s.t.}:\hspace{.25cm}
	\eqref{cons userassociation dn1}\text{-}\eqref{cons min rate user up},\\&
	\hspace{1cm}\Phi^\text{ISP}_\text{tot}\ge\Phi^\text{ISP}_0,\\&
			\hspace{1cm}\Phi^\text{User}_\text{tot}\ge \Phi^\text{User}_0,\\&
						\hspace{1cm}\Phi^\text{InP}_{i'}\ge \Phi^\text{EInP}_0, i'\in \mathcal{I}/i,\\&
	\hspace{1cm}\Phi^\text{SDO}_\text{tot}\ge \Phi^\text{SDO}_0.
	\end{align}
\end{subequations}
\subsection{ISP Optimization Problem}
Each ISP maximizes its utility function by considering minimum utilities $\Phi^\text{InP}_0$, $\Phi^\text{User}_0$ and $\Phi^\text{SDO}_0$ for InPs, users and SDO, respectively. Moreover, for other ISPs, each ISP considers minimum utility $\Phi^\text{EISP}_0$.  The corresponding optimization problem is formulated as:
\begin{subequations}\label{MOO0122}
	\begin{align}\label{MOO01a22}
	&\max_{\boldsymbol{\rho},\boldsymbol{p},\boldsymbol{L},\boldsymbol{\alpha}}
	\Phi^\text{ISP}_v,
	\\&\nonumber
	\text{s.t.}:\hspace{.25cm}
	\eqref{cons userassociation dn1}\text{-}\eqref{cons min rate user up},\\&
	\hspace{1cm}\Phi^\text{InP}_\text{tot}\ge \Phi^\text{InP}_0,\\&
			\hspace{1cm}\Phi^\text{User}_\text{tot}\ge \Phi^\text{User}_0,\\&
			\hspace{1cm}\Phi^\text{ISP}_{v'}\ge \Phi^\text{EISP}_0,\forall v'\in \mathcal{V}/v,\\&
	\hspace{1cm}\Phi^\text{SDO}_\text{tot}\ge \Phi^\text{SDO}_0.
	\end{align}
\end{subequations}
\subsection{SDO Optimization Problem}
 SDO maximizes its utility function by considering minimum utilities $\Phi^\text{InP}_0$, $\Phi^\text{User}_0$ and $\Phi^\text{ISP}_0$ for InPs, users, and ISPs, respectively. The corresponding optimization problem is formulated as:
\begin{subequations}\label{MOO0123}
	\begin{align}\label{MOO01a23}
	&\max_{\boldsymbol{\rho},\boldsymbol{p},\boldsymbol{L},\boldsymbol{\alpha}}
	\Phi^\text{SDO}_\text{tot},
	\\&\nonumber
	\text{s.t.}:\hspace{.25cm}
	\eqref{cons userassociation dn1}\text{-}\eqref{cons min rate user up},\\&
	\hspace{1cm}\Phi^\text{InP}_\text{tot}\ge \Phi^\text{InP}_0,\\&
		\hspace{1cm}\Phi^\text{User}_\text{tot}\ge \Phi^\text{User}_0,\\&
	\hspace{1cm}\Phi^\text{ISP}_\text{tot}\ge \Phi^\text{ISP}_0.
	\end{align}
\end{subequations}


\subsection{Central Unit Optimization Problem}
As we mentioned,  central unit by considering the reported prices, solves a joint power and codebook allocation problem. Therefore, optimization problem which should be solved by the central unit is presented as follows: 

	\begin{align}\label{MOO0101}
		\max_{\boldsymbol{\rho},\boldsymbol{p},\boldsymbol{\alpha} }
		\{\Phi^\text{ISP}_\text{tot},
		\Phi^\text{InP}_\text{tot}, 
		\Phi^\text{Sens}_\text{tot},
		\Phi^\text{User}_\text{tot}\},
		\text{s.t.}:\hspace{.25cm}
		\eqref{cons userassociation dn1}\text{-}\eqref{cons min rate user up}.
	\end{align}

\section{Solution}\label{Section Solution}
\subsection{Solution of the Weight-One  Approach }
In order to solve the optimization problem \eqref{problem main}, the alternating algorithm is used \cite{hooke1961direct}. Based on the alternating method, in each iteration, each set of variables are calculated assuming other variable sets are fixed.  

The main solution steps are presented in Algorithm \ref{algorithm-1}.
\begin{algorithm}
	\caption{ITERATIVE RESOURCE ALLOCATION ALGORITHM }
	\label{algorithm-1}
			\begin{itemize}
\item	STEP1: Initialization:
		\begin{itemize}
			\item
	 Set $t=0$ (iteration number),
            \item
	 Find $\boldsymbol{\alpha}(0)$, $\mathbf{P}(0)$, $\mathbf{L}(0)$, and $\boldsymbol{\rho}(0)$
		\end{itemize}
\item	STEP2:
	\begin{itemize}
		\item
		Set $\mathbf{P}=\mathbf{P}(t)$,  $\boldsymbol{\alpha}=\boldsymbol{\alpha}(t)$, and $\boldsymbol{\rho}=\boldsymbol{\rho}(t)$,
		\item
		Solve the optimization problem with variable $\mathbf{L}$,
		\item
		Set the  result of optimization problem solution to $\mathbf{L}(t+1)$,
	\end{itemize}
\item	STEP3:
	\begin{itemize}
		\item
		Set $\mathbf{P}=\mathbf{P}(t)$, $\mathbf{L}=\mathbf{L}(t+1)$, and $\boldsymbol{\rho}=\boldsymbol{\rho}(t)$,
		\item
	  Solve the optimization problem with variable $\boldsymbol{\alpha}$,
	    \item
	  Set the  result of optimization problem solution to $\boldsymbol{\alpha}(t+1)$,
	\end{itemize}
\item	STEP4:
\begin{itemize}
	\item
	Set $\mathbf{L}=\mathbf{L}(t+1)$, $\boldsymbol{\alpha}=\boldsymbol{\alpha}(t+1)$, and $\boldsymbol{\rho}=\boldsymbol{\rho}(t)$,
	\item
	Solve the optimization problem with variable $\mathbf{P}$,
	\item
	Set the  result of optimization problem solution to $\mathbf{P}(t+1)$,
\end{itemize}
\item	STEP5:
\begin{itemize}
	\item
	Set $\mathbf{L}=\mathbf{L}(t+1)$, $\boldsymbol{\alpha}=\boldsymbol{\alpha}(t+1)$, and $\mathbf{P}=\mathbf{P}(t+1)$,
	\item
	Solve the optimization problem with variable $\boldsymbol{\rho}$,
	\item
	Set the  result of optimization problem solution to $\boldsymbol{\rho}(t+1)$,
\end{itemize}
\item	STEP6:
	\begin{itemize}
	\item If convergence\\
	stop and return $\boldsymbol{\alpha}$, $\mathbf{P}$, $\mathbf{L}$, and $\boldsymbol{\rho}$ as the suboptimal solution,
	\item Else\\
	set $t=t+1$ and go back to STEP 2,
	\item Output:\,\,
	Suboptimal  value of $\boldsymbol{\alpha}$, $\mathbf{P}$, $\mathbf{L}$, and $\boldsymbol{\rho}$.
	\end{itemize}
	\end{itemize}
\end{algorithm}
In Step 2, the problem of
finding $\boldsymbol{L}$ is solved by assuming the other variables being fixed. This problem  is a non-constrained linear programming which can be solved by   using the  CVX toolbox \cite{11002233}.
 The optimization problem with  variable $\boldsymbol{L}$  is formulated as follows:
\begin{align}\label{obf problem alpha}
&\max_{ \boldsymbol{L} }
\sum_{v \in \mathcal{V}} \omega^\text{ISP}_v \Phi^\text{ISP}_v + \sum_{u \in \mathcal{U}} \omega^\text{User}_u \Phi^\text{User}_u\\& \nonumber + \sum_{i \in \mathcal{I}} \omega^\text{InP}_i \Phi^\text{InP}_i + \sum_{i \in \mathcal{I}} \sum_{b_i \in \mathcal{B}_i} \sum_{s_{b_i} \in \mathcal{S}_{b_i}} \omega^\text{Sens}_{s_{b_i}} \Phi^\text{Sens}_{s_{b_i}}.
\end{align}
In Step 3, with assumption of other optimization variables being fixed, the problem of finding $\boldsymbol{\alpha}$ is solved.
By using the epigraph technique and relaxing the variable  $\boldsymbol{\alpha}$ to have the continuous value between $0$ and $1$, the optimization problem \eqref{obf problem alpha} is reformulated as:
\begin{subequations}\label{problem alpha epi}
	\begin{align}\label{obf problem alpha epi}
	& \max_{ \boldsymbol{\alpha} , \boldsymbol{\delta} , \boldsymbol{\beta} }
    \sum_{i \in \mathcal{I}} \sum_{b_i \in \mathcal{B}_i} \sum_{s_{b_i} \in \mathcal{S}_{b_i}} \omega^\text{Sens}_{s_{b_i}} \bigg( \sum\limits_{v \in \mathcal{V}} \delta_{v,s_{b_i}} L^\text{Sens,Data}_{v,s_{b_i}} + \sum\limits_{v \in \mathcal{V}} \sum\limits_{c'_i \in \mathcal{UC}_i} \delta_{v,s_{b_i}}  r^{c'_i}_{b_i,s_{b_i}} L^\text{Up,Rate}_{s_{b_i}} - \beta_{s_{b_i}} C^\text{Sens,Reserv}_{s_{b_i}} \bigg) + \nonumber   \\\nonumber
    &\sum_{v \in \mathcal{V}} \omega^\text{ISP}_v \bigg(
    \sum_{ u \in \mathcal{K}_v} q \log( 1+\frac{\sum\limits_{i=1}^{I} \sum\limits_{b_i \in \mathcal{B}_i} \sum\limits_{s_{b_i} \in \mathcal{S}_{b_i}} \alpha_{s_{b_i},u}} { \sum\limits_{i=1}^{I} \sum\limits_{b_i \in \mathcal{B}_i} S_{b_i}}) L^\text{Reserv,User}_{v,u} - \sum_{s_{b_i} \in \mathcal{S}_{b_i}} \delta_{v,s_{b_i}} L^\text{Sens,Data}_{v,s_{b_i}} -     \nonumber   \\\nonumber
    &  \sum_{i \in \mathcal{I}} \sum_{b_i \in \mathcal{B}_i} \sum_{s_{b_i} \in \mathcal{S}_{b_i}} \delta_{v,s_{b_i}}  r^{c'_i}_{b_i,s_{b_i}} L^\text{Up,Rate}_{s_{b_i}}
    \bigg)    +  \sum_{v \in \mathcal{V}} \sum_{u \in \mathcal{K}_v} \omega^\text{User}_u \bigg(
    q \log( 1+\frac{\sum\limits_{i=1}^{I} \sum\limits_{b_i \in \mathcal{B}_i} \sum\limits_{s_{b_i} \in \mathcal{S}_{b_i}} \alpha_{s_{b_i},u}} { \sum\limits_{i=1}^{I} \sum\limits_{b_i \in \mathcal{B}_i} S_{b_i}})\\& \nonumber (C^\text{Reserv,User}_{u}-L^\text{Reserv,User}_{v,u})
    \bigg),
    \\
    & \text{s.t.}~~~
	\delta_{v,s_{b_i}} \leq \sum\limits_{u \in \mathcal{K}_v} \alpha_{s_{b_i},u},    \\
    & ~~~~~~ \delta_{v,s_{b_i}} \leq 1,  \\
    & ~~~~~~   \beta_{s_{b_i}} \leq \sum\limits_{u \in \mathcal{U}} \alpha_{s_{b_i},u},    \\
    & ~~~~~~   \beta_{s_{b_i}} \leq 1,     \\
    & ~~~~~~   0 \leq \alpha_{s_{b_i},u} \leq 1,
	\end{align}
\end{subequations}
where $\boldsymbol{\delta}=[\delta_{v,s_{b_i}}]$ and  $\boldsymbol{\beta}=[\beta_{s_{b_i}}]$ are the auxiliary variables corresponding to the epigraph algorithm. The optimization  problem \eqref{problem alpha epi} is  convex  which can be solved by applying the CVX toolbox.
 The aim of Step 4 is finding $\boldsymbol{p}$. The corresponding optimization problem is formulated as:
\begin{subequations}\label{problem power}
	\begin{align}\label{obf problem power}
	& \max_{ \boldsymbol{p} }
    \sum_{v \in \mathcal{V}} \omega^\text{ISP}_v \bigg(
    \sum_{ i \in \mathcal{I}} \sum_{ b_i \in \mathcal{B}_i} \sum_{ u \in \mathcal{K}_v} \sum_{c_i \in \mathcal{DC}_i} L^\text{Dn,Rate}_v W_\text{S} r^{c_i}_{b_i,u} \\& \nonumber- \sum_{ i \in \mathcal{I}} \sum_{ b_i \in \mathcal{B}_i} \sum_{ u \in \mathcal{K}_v} \sum_{c_i \in \mathcal{DC}_i} L^\text{Power,BS}_{b_i} \rho^{c_i}_{b_i,u} p^{c_i}_{b_i,u} -    \nonumber   \\\nonumber
    & \sum_{i \in \mathcal{I}} \sum_{b_i \in \mathcal{B}_i} \sum_{s_{b_i} \in \mathcal{S}_{b_i}} \min \{ \sum\limits_{u \in \mathcal{K}_v} \alpha_{s_{b_i},u} , 1 \} r^{c'_i}_{b_i,s_{b_i}} L^\text{Up,Rate}_{s_{b_i}}
    \bigg) -    \nonumber   \\\nonumber
    & \sum_{v \in \mathcal{V}} \sum_{u \in \mathcal{K}_v} \omega^\text{User}_u \bigg(
    \sum_{ i \in \mathcal{I}} \sum_{ b_i \in \mathcal{B}_i} \sum_{c_i \in \mathcal{DC}_i} L^\text{Dn,Rate}_v W_\text{S} r^{c_i}_{b_i,u}
    \bigg) +\\& \nonumber
    \sum_{i \in \mathcal{I}} \omega^\text{InP}_i \bigg(
    \sum_{ b_i \in \mathcal{B}_i} \sum_{ u \in \mathcal{U}} \sum_{c_i \in \mathcal{DC}_i} L^\text{Power,BS}_{b_i} \rho^{c_i}_{b_i,u} p^{c_i}_{b_i,u} \bigg) + \nonumber   \\
    &\nonumber \sum_{i \in \mathcal{I}} \sum_{b_i \in \mathcal{B}_i} \sum_{s_{b_i} \in \mathcal{S}_{b_i}} \omega^\text{Sens}_{s_{b_i}} \bigg(
    \sum\limits_{v \in \mathcal{V}} \sum\limits_{c'_i \in \mathcal{UC}_i} \min \{ \sum\limits_{u \in \mathcal{K}_v} \alpha_{s_{b_i},u} , 1 \} \\& \nonumber r^{c'_i}_{b_i,s_{b_i}} L^\text{Up,Rate}_{s_{b_i}} -
    \sum\limits_{c'_i \in \mathcal{UC}_i} \rho^{c'_i}_{b_i,s_{b_i}} p^{c'_i}_{b_i,s_{b_i}} C^\text{Power,Sup}
    \bigg),
    \\
    & \text{s.t.}\hspace{.1cm}
	\eqref{cons max power BS}\text{-}\eqref{cons min rate user up}.    \nonumber
	\end{align}
\end{subequations}
Due to the non-convex rate function in uplink and downlink transmissions, the optimization problem \eqref{problem power} is  non-convex. To tackle the non-convexity issue of the considered problem, a successive convex approximation (SCA) approach with difference of two concave functions (D.C.) approximation method is used.
In order to apply  this  method, at first the downlink  rate function is written as
$
r^{c_i}_{b_i,u} = f^{c_i}_{b_i,u} - g^{c_i}_{b_i,u},
$
where
\begin{align}\label{f downlink BS}
&f^{c_i}_{b_i,u} = \log_2 (\rho^{c_i}_{b_i,u}\sum_{n\in \mathcal{N}_i}q^{c_i}_{n}\lambda^{b_i}_{n,c_i}p^{c_i}_{b_i,u}|h^{n}_{b_i,u}|^2+I^{c_i}_{b_i,u}+(\sigma^{c_i}_{b_i,u})^2 ),
\end{align}
\begin{align}\label{g downlink BS}
g^{c_i}_{b_i,u} = \log_2 ( I^{c_i}_{b_i,u}+(\sigma^{c_i}_{b_i,u})^2).
\end{align}
By applying the D.C. approximation, we have
\begin{align}\label{g Downlink BS approximated}
&g^{c_i}_{b_i,u}(\boldsymbol{p}^{\text{BS},t}) \approx  g^{c_i}_{b_i,u}(\boldsymbol{p}^{\text{BS},t-1}) +
\nabla g^{c_i}_{b_i,u}(\boldsymbol{p}^{\text{BS},t-1})
(\boldsymbol{p}^{\text{BS},t} - \boldsymbol{p}^{\text{BS},t-1}),
\end{align}
where
\begin{align}\label{g gradian Downlink BS}
&\nabla g^{c_i}_{b_i,u} (\boldsymbol{p}^\text{BS})  =
\left\{
  \begin{array}{ll}
    0, & \hbox{$\forall i' \neq i$}, \\
    \frac{\sum_{n\in \mathcal{N}_i}\rho^{c_i}_{b'_i,u'}q^{c_i}_{n}\lambda^{b'_i}_{n,c_i}|h^{n}_{b'_i,u}|^2}  {\ln(2) \big( I^{c_i}_{b_i,u}+(\sigma^{c_i}_{b_i,u})^2 \big)}, & \hbox{$\forall u' \neq u, b'_i \neq b_i$}.
  \end{array}
\right.
\end{align}

The uplink data rate function is written as
\begin{align}\label{rate uplink BS MENHA}
r^{c'_i}_{b_i,s_{b_i}} = f'^{c'_i}_{b_i,s_{b_i}} - g'^{c'_i}_{b_i,s_{b_i}},
\end{align}
where
\begin{align}\label{f uplink BS}
&f^{c'_i}_{b_i,s_{b_i}} = \log_2 ( \rho'^{c'_i}_{b_i,s_{b_i}} \sum_{m\in \mathcal{M}_i}q'^{m}_{b_i,s_{b_i}} \lambda'^{b_i}_{n,c'_i}p'^{c'_i}_{b_i,s_{b_i}}|h^{m}_{b_i,s_{b_i}}|^2+ I'^{c'_i}_{b_i,s_{b_i}}+(\sigma'^{c'_i}_{b_i,s_{b_i}})^2),
\end{align}
\begin{align}\label{g uplink BS}
g^{c'_i}_{b_i,s_{b_i}} = \log_2 (I'^{c'_i}_{b_i,s_{b_i}}+(\sigma'^{c'_i}_{b_i,s_{b_i}})^2 ).
\end{align}
By applying the D.C. approximation, we have
\begin{align}\label{g uplink BS approximated}
&g^{c'_i}_{b_i,s_{b_i}}(\boldsymbol{p}^{\text{Sens},t}) \approx  g^{c'_i}_{b_i,s_{b_i}}(\boldsymbol{p}^{\text{Sens},t-1}) +
\nabla g^{c'_i}_{b_i,s_{b_i}}(\boldsymbol{p}^{\text{Sens},t-1})
(\boldsymbol{p}^{\text{Sens},t} - \boldsymbol{p}^{\text{Sens},t-1}).
\end{align}
where
\begin{align}\label{g gradian uplink BS}
&\nabla g^{c'_i}_{b_i,s_{b_i}} (\boldsymbol{p}^\text{Sens})  =
\left\{
  \begin{array}{ll}
    0, & \hbox{$\forall i' \neq i$}, \\
    \frac{\sum_{m\in \mathcal{M}_i}\rho'^{c'_i}_{b'_i,s'_{b'_i}}q^{c'_i}_{m}\lambda'^{b'_i}_{m,c'_i}|h^m_{b_i,s'_{b'_i}}|^2 }{ I'^{c'_i}_{b_i,s_{b_i}}+(\sigma'^{c'_i}_{b_i,s_{b_i}})^2}. & \hbox{$b'_i \neq b_i$}.
  \end{array}
\right.
\end{align}
By applying the D.C. approximation, the optimization problem \eqref{problem power} is approximated by a convex function which can be solved by CVX toolbox.
In Step 5, the problem of finding  $\boldsymbol{\rho}$ is solved which is an integer non-linear optimization problem. To solve it, we initially  relax the integer variables to  continuous values between zero and one. The same steps that are applied in the previous step are applied here as well.

\subsection{Solution of the Max-Min Approach }
To solve the max-min approach  optimization problem,  at first we apply the  epigraph method as:
\begin{subequations}\label{MOO012}
	\begin{align}\label{MOO01a2}
	&\max_{\boldsymbol{\rho},\boldsymbol{p},\boldsymbol{L},\boldsymbol{\alpha},t }
	t+
	\Phi^\text{User}_\text{tot},
	\\&\nonumber
	\text{s.t.}:\hspace{.25cm}
	\eqref{cons userassociation dn1}\text{-}\eqref{cons min rate user up},\\&
	\hspace{1cm}\Phi^\text{ISP}_\text{tot}\ge t,\\&
	\hspace{1cm}\Phi^\text{InP}_\text{tot}\ge t,\\&
\hspace{1cm}	\Phi^\text{Sens}_\text{tot}\ge t,\nonumber
	\end{align}
\end{subequations}
where $t$ is an auxiliary variable. 
Then, we continue similar to the weight-one approach algorithm.
%

\subsection{Solution of the Conventional Approach Problems }
Each of the presented optimization problem in the conventional approach can be solved by the iterative algorithm shown in Algorithm \ref{algorithm-1}.

In order to achieve and optimal solution of for the proposed optimization problem,  
 the monotonic optimization approach can be applied \cite{7472351, 7862919} which needs some alterations in the objective and
 constraints to convert the original problem into the standard form of the monotonic optimization
 problems. However, in this paper, due to the space limitation and huge computational complexity,
 we omit this approach in the paper.

\section{Convergence of the Proposed Solution Algorithm}\label{Convergence of the Proposed Solution Algorithm}
The alternating method   exploits an iterative algorithm in which in each iteration each set of variables is calculated by supposing that the other variable sets are fixed and process is continued until convergence. The necessary and sufficient  conditions to ensure the algorithm convergence is that in each iteration the objective function increases or stays unaltered compared to the previous iteration \cite{4752799,5771610}. In our solution, we have 
\begin{align}\label{prty}
&\dots \le U(\boldsymbol{\rho}^{t},\boldsymbol{p}^{t},\boldsymbol{L}^{t},\boldsymbol{\alpha}^{t})\stackrel{a}{\le} U(\boldsymbol{\rho}^{t},\boldsymbol{p}^{t},\boldsymbol{L}^{t+1},\boldsymbol{\alpha}^{t})\stackrel{b}{\le}\\\nonumber& U(\boldsymbol{\rho}^{t},\boldsymbol{p}^{t},\boldsymbol{L}^{t+1},\boldsymbol{\alpha}^{t+1})\stackrel{c}{\le}
 U(\boldsymbol{\rho}^{t},\boldsymbol{p}^{t+1},\boldsymbol{L}^{t+1},\boldsymbol{\alpha}^{t+1})\stackrel{d}{\le}
 \\\nonumber&
U(\boldsymbol{\rho}^{t+1},\boldsymbol{p}^{t+1},\boldsymbol{L}^{t+1},\boldsymbol{\alpha}^{t+1})\le\dots
\end{align}
which indicates that after each iteration, the objective function increases or stays unaltered compared to the previous iteration. 

Inequality (a) in  \eqref{prty} follows from the fact that optimization problem with variables $\boldsymbol{L}$ and constant $\boldsymbol{\rho},\boldsymbol{p}$ and $\boldsymbol{\alpha}$ ($\boldsymbol{\rho}=\boldsymbol{\rho}^{t},\boldsymbol{p}=\boldsymbol{p}^{t},\boldsymbol{\alpha}=\boldsymbol{\alpha}^{t}$) is a linear program whose worst solution is $\boldsymbol{L}=\boldsymbol{L}^t$, therefore, based on the worst solution, we have $\boldsymbol{L}^{t+1}=\boldsymbol{L}^t$. Consequently, we have  $U(\boldsymbol{\rho}^{t},\boldsymbol{p}^{t},\boldsymbol{L}^{t},\boldsymbol{\alpha}^{t})\le^1 U(\boldsymbol{\rho}^{t},\boldsymbol{p}^{t},\boldsymbol{L}^{t+1},\boldsymbol{\alpha}^{t})$. For inequalities (b), (c) and (d), the same argument as  for inequality (a) can be used. Since for a finite set of transmit powers and subcarrier assignment, the summation of utilities is bounded, the procedure must converge.

\section{ Computational Complexity}\label{section Computational Complexity}
In this section, we investigate the computational complexity of the proposed methods. 
As mentioned, we used   Algorithm \ref{algorithm-1} to solve problem \eqref{MOO010} in three different approaches. The Algorithm has four stages which determine $ \boldsymbol{\rho} $, $ \boldsymbol{p} $, $ \boldsymbol{\alpha} $ and $ \boldsymbol{L} $. In all employed methods, we use DC approximation to solve subcarrier and power allocation problems. The main computational
complexity for the DC approximation comes from solving the problem via CVX which applies interior point method. Generally, the number of required iterations of interior point method is $\dfrac{\text{log}((\Delta)/t^0\varrho)}{\text{log}(\xi_e)}$ where $ \Delta $ is the total number of constraints, $ t^0 $ is initial point to approximated the accuracy of
interior point method, $ \varrho $ is the stopping criterion and $ \xi_e $ is used to update the accuracy
of interior point method.
 CVX uses interior point method for variables $ \boldsymbol{\alpha} $ and $ \boldsymbol{L} $ as $ \boldsymbol{\rho} $ and $ \boldsymbol{p} $. Hence, their required iterations are computed similar to $ \boldsymbol{\rho} $ and $ \boldsymbol{p} $. Number of iterations in all proposed methods differ in $ \Delta$. In Weight-One approach, total number of constraints due to \eqref{cons userassociation dn1}\text{-}\eqref{cons min rate user up} equals $ N_{\boldsymbol{\rho}}= U\sum_{ i \in \mathcal{I}}C_i\sum_{ i \in \mathcal{I},i'\neq i}C_{i'}\sum_{ i \in \mathcal{I}}B_i\sum_{ i \in \mathcal{I},i'\neq i}B_{i'} + U\sum_{ i \in \mathcal{I}}C_i^2B_{i}(B_{i}-1)+\sum_{ i \in \mathcal{I}}N_i+\sum_{ i \in \mathcal{I}}M_i+\sum_{ i \in \mathcal{I}}B_i+2S+U$ for subcarrier allocation and $ N_{\boldsymbol{p}}= \sum_{ i \in \mathcal{I}}B_i+2S+U$ for power allocation stages. Number of constraints of \eqref{problem alpha epi} and \eqref{obf problem alpha} determine $ \Delta $ of $ \boldsymbol{\alpha} $ and $ \boldsymbol{L} $, respectively. In Max-Min approach, number of constrants of \eqref{MOO012} is calculated for all variables. In conventional approach, $ \Delta $ for variable $ \boldsymbol{L} $ is calculated based on 
\eqref{MOO0121}\textbf{-}\eqref{MOO0123} and for other variables $ \Delta $ is determined from \eqref{MOO0101}.
 All the number of constraints of these approaches are summarized in Table \ref{table2}.
\begin{table}[h] \centering  
\caption{Computational Complexity of Proposed Methods}
\begin{tabular}{ccc}\label{table2}
 Approach & Variable &  $ \Delta $ \\ 
\hline $  $ & $ \boldsymbol{\rho} $ &   $ N_{\boldsymbol{\rho}} $  \\
 \cline{2-3}
 Weight-One & $ \boldsymbol{p} $ & $ N_{\boldsymbol{p}} $\\
  \cline{2-3}
   $  $ & $ \boldsymbol{\alpha} $ & $ S(2V+U+2) $\\
   \cline{2-3}
    $  $ & $ \boldsymbol{L} $ & $ 0 $\\
      \cline{2-3}
\hline \hline $  $ & $ \boldsymbol{\rho} $ &  $  N_{\boldsymbol{\rho}}+3 $  \\
 \cline{2-3}
 Max-Min & $ \boldsymbol{p} $ & $ N_{\boldsymbol{p}}+3 $\\
   \cline{2-3}
    $  $ & $ \boldsymbol{\alpha} $ & $ S(2V+U+2)+2 $\\
    \cline{2-3}
     $  $ & $ \boldsymbol{L} $ & $ 3 $\\
       \cline{2-3}
      \hline \hline  $  $ & $ \boldsymbol{\rho} $ &  $  N_{\boldsymbol{\rho}}+I+V+9 $  \\
       \cline{2-3}
       Conventional & $ \boldsymbol{p} $ & $ N_{\boldsymbol{p}}+I+V+9 $\\
         \cline{2-3}
          $  $ & $ \boldsymbol{\alpha} $ & $ S(2V+U+2)+V+7 $\\
          \cline{2-3}
           $  $ & $ \boldsymbol{L} $ & $ I+V+9 $\\
             \cline{2-3}
                   \hline 
\end{tabular} 
\end{table}

\section{Numerical Results and Discussion}\label{section simulation results}
In order to evaluate the performance of the proposed end to end pricing and radio resource allocation approach, we first depict the utility of each of the players versus various parameters. Moreover, we compare the performance of the proposed approach to the  conventional  one. It is worth to note that due to the large number of variables and run-time limitation, without loss of generality,  we present the numerical results for a few number of   IoT users, base stations, and sensors.  
\subsection{Parameters}

The considered parameters for numerical results are summarized as follows: $P_{1_i}^{\text{max}}=50$ (Watts) for all $i\in \mathcal{I}$, $P_{b_i}^{\text{max}}=1$ (Watts) for all $i\in \mathcal{I}$ and $b\in \{2,\dots,B\}$, $I=2$, $B=2$, $V=2$, $R^\text{min}_{\text{Up},s_{b_i}}=0.01$ (bps/Hz) for all $ b_i \in \mathcal{B}_i, s_{b_i} \in \mathcal{S}_{b_i},$  and $R^\text{min}_{\text{Dn},v}=0.1$ (bps/Hz) for all $v \in \mathcal{V}, v \in \mathcal{K}_v$, $h^{n}_{b_i,u}=x^{n}_{b_i,u}(D_{b_i,u})^{\xi}$ where $\xi$ indicates the path loss
exponent and $\xi = -3$, 
 $x^{n}_{b_i,u}$ represents the Rayleigh
fading, and $D_{b_i,u}$ demonstrates the distance between user $u$
and BS $b_i$. Moreover, to scale the price of the power, data of sensor, and user reservation of player, we define a parameter as scale of player with value  $SF=10^5$. By defining this scale of player, the maximum and minimum prices of each of the parameters is defined as follows: $0\preceq\boldsymbol{L}^\text{Power,BS}\preceq SF\times L_{\text{max}}$, $0\preceq\boldsymbol{L}^\text{BW}\preceq L_{\text{max}}$, $0\preceq\boldsymbol{L}^\text{Sens,Data}\preceq SF\times L_{\text{max}}$,
$0\preceq \boldsymbol{L}^\text{Up,Rate}\preceq L_{\text{max}}$, $0\preceq \boldsymbol{L}^\text{Dn,Rate}\preceq L_{\text{max}}$ and 
$0\preceq \boldsymbol{L}^\text{Reserv,User}\preceq SF\times L_{\text{max}}$ where $\preceq$ denotes vector inequality or componentwise inequality. 
Moreover, there are some common settings in the most of the results which are as follows: $ S_{b_i}=3 $ and the number of subcarriers for both of UL and DL transmission are set to 4. Furthermore, we assume that each ISP serves 4 users. 
\subsection{Results}  
\begin{figure}
	\centering
	\subfigure{
		{\includegraphics[width=.47\textwidth]{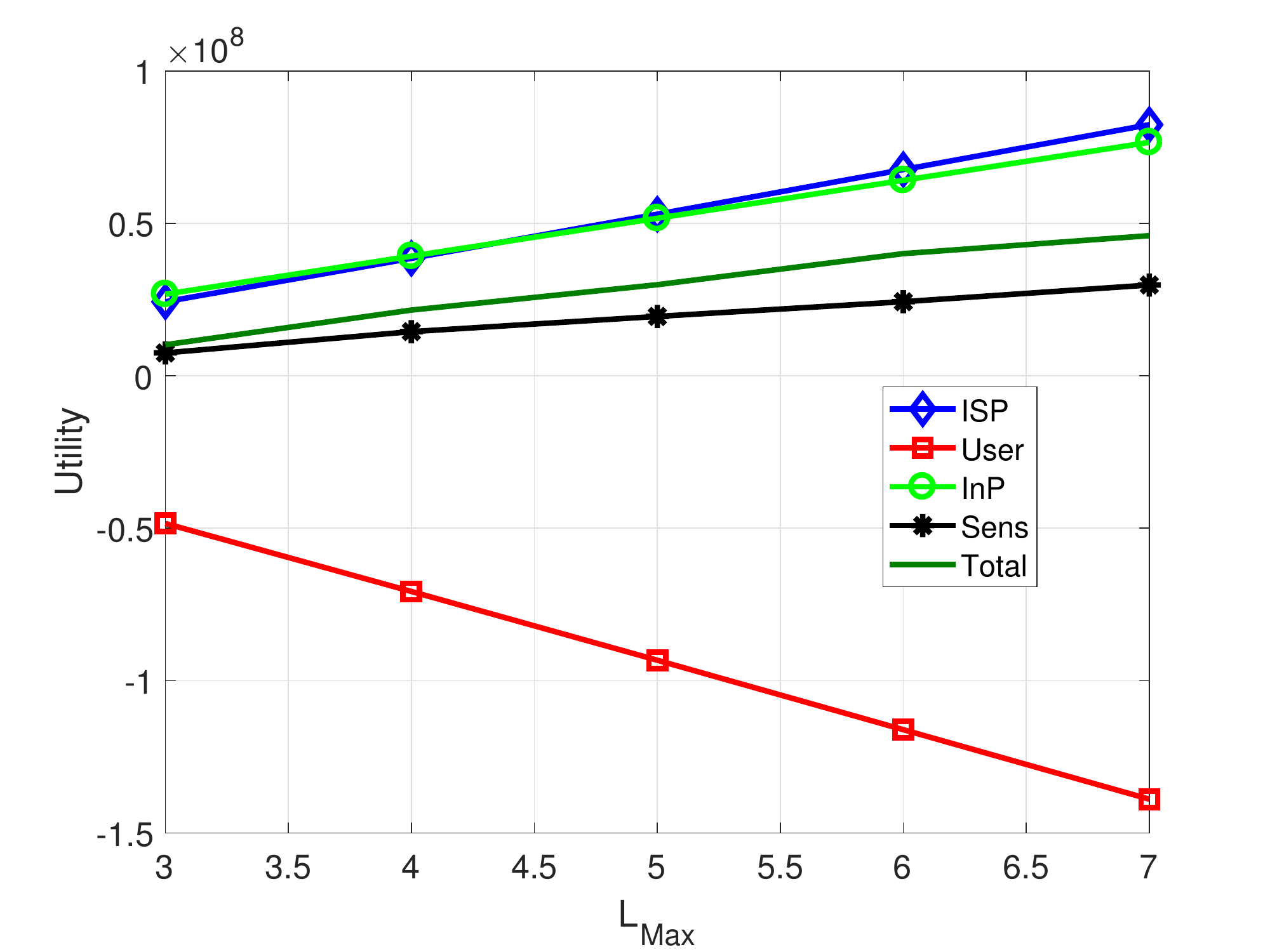}}
	}
	\subfigure{
		{\includegraphics[width=.44\textwidth]{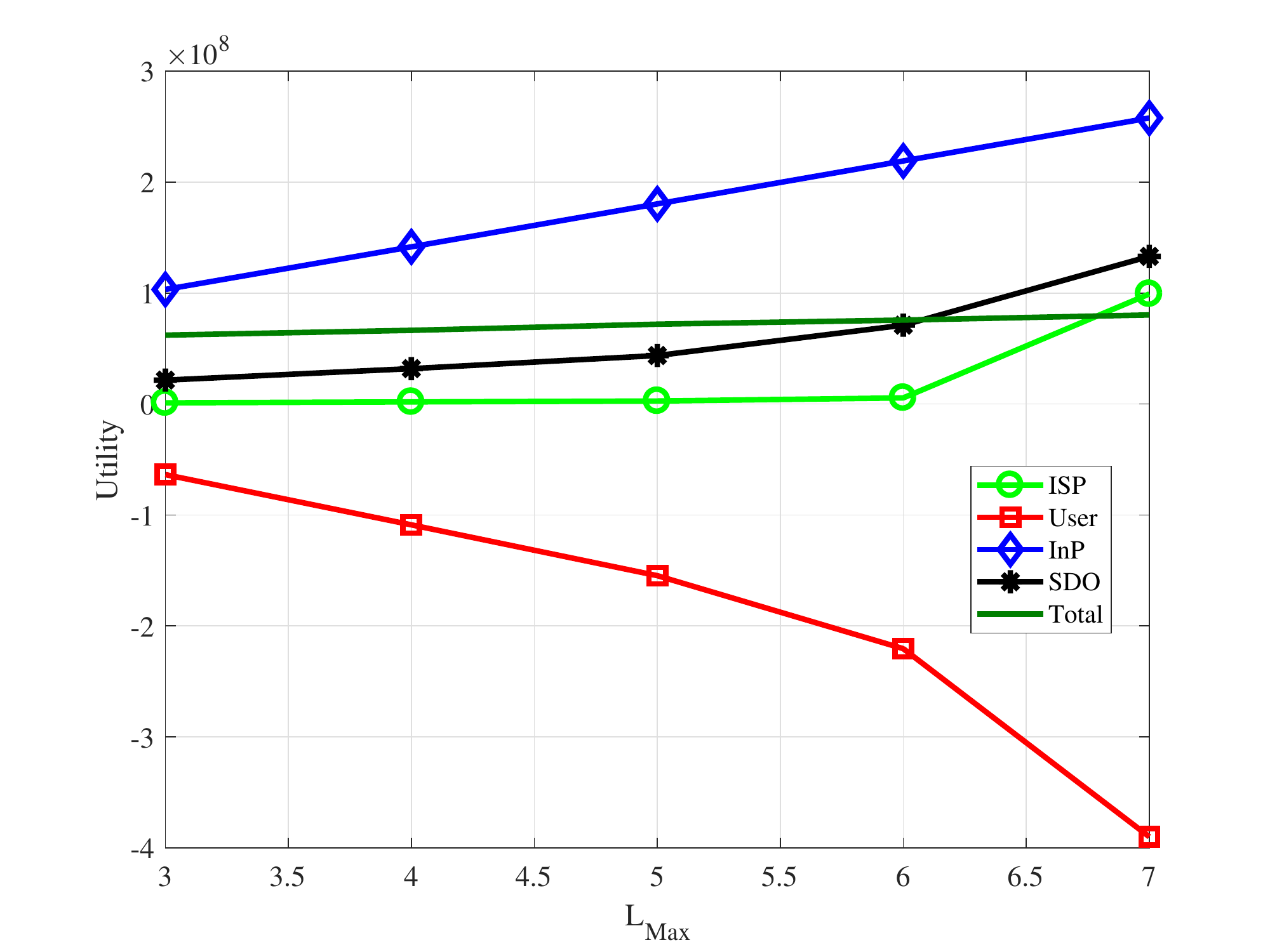}}
	}
	\caption{Left:Utility of each player in the max-min approach versus different values of $L_\text{max}$. Right: Utility of each player in the  weight-one approach versus different values of $L_{\text{max}}$.}
	\label{max_min_lmax}
\end{figure}

\begin{figure}
	\centering
	\includegraphics[width=.45\textwidth]{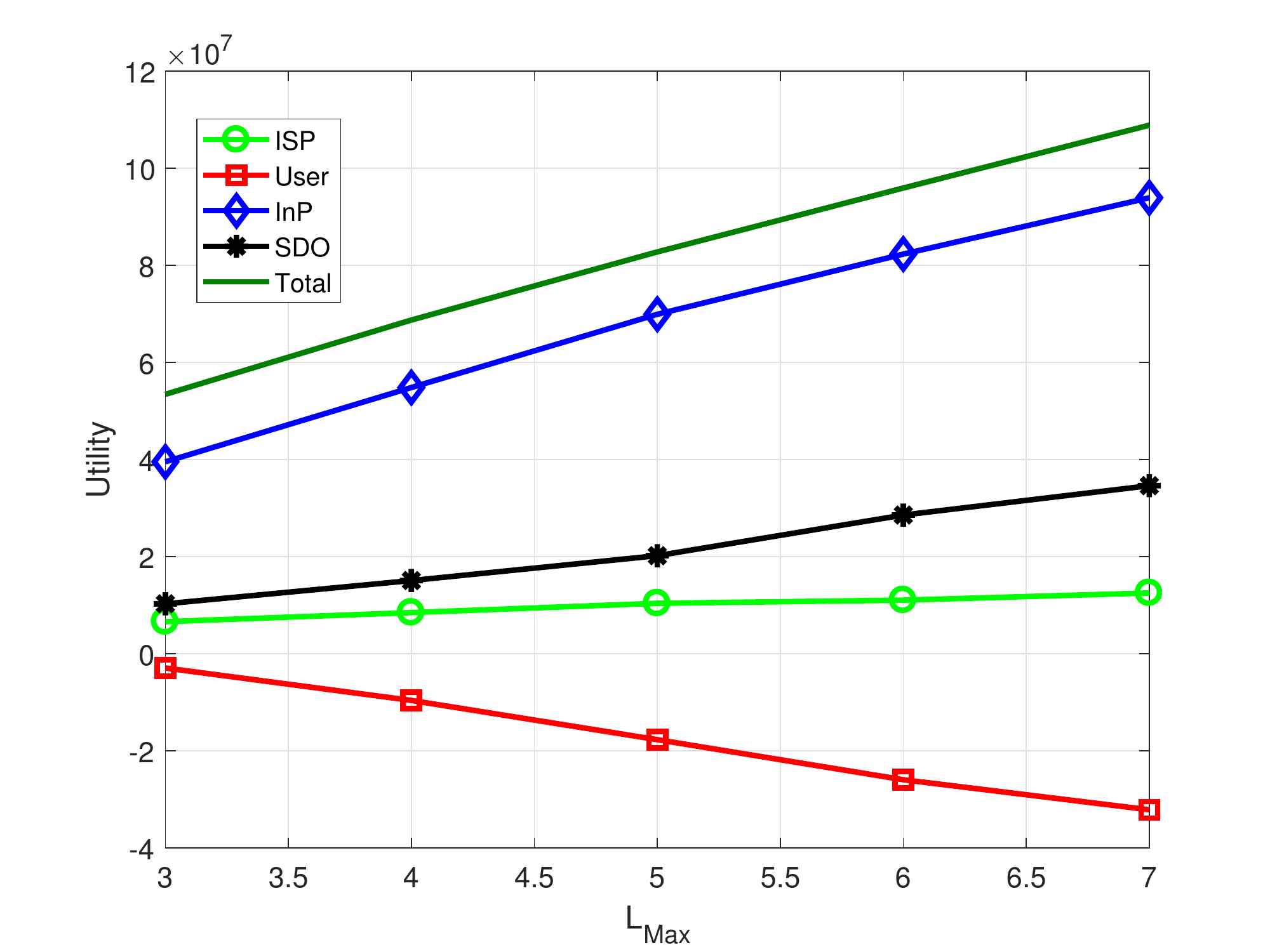}
	\caption{Utility of each player in the  conventional approach  versus different values of $L_{\text{max}}$.}
	\label{pmaxdiss}
\end{figure}

 Figs. \ref{max_min_lmax}-Left, \ref{max_min_lmax}-Right, \ref{pmaxdiss}, respectively, depict the utility of all players in the max-min, weight-one, and conventional  approach versus different values of $L_{\text{max}}$ where the number of sensors is set to 3. From these figures we can see that by increasing the utility (revenue) of InPs, ISPs, and SDO, the total utility of the users is decreased. This is due to the fact that users are the  end consumer of the network and the other players income comes from their payments. Moreover, as can be seen,  the utility of InPs is more than that of the other players. This is because InPs sell their resources to  SDO and ISPs, simultaneously (see the details of pricing model of InPs in n in Fig. \ref{IoT pricing scheme}). 
%

\begin{figure}
	\centering
	\subfigure{
		{\includegraphics[width=.45\textwidth]{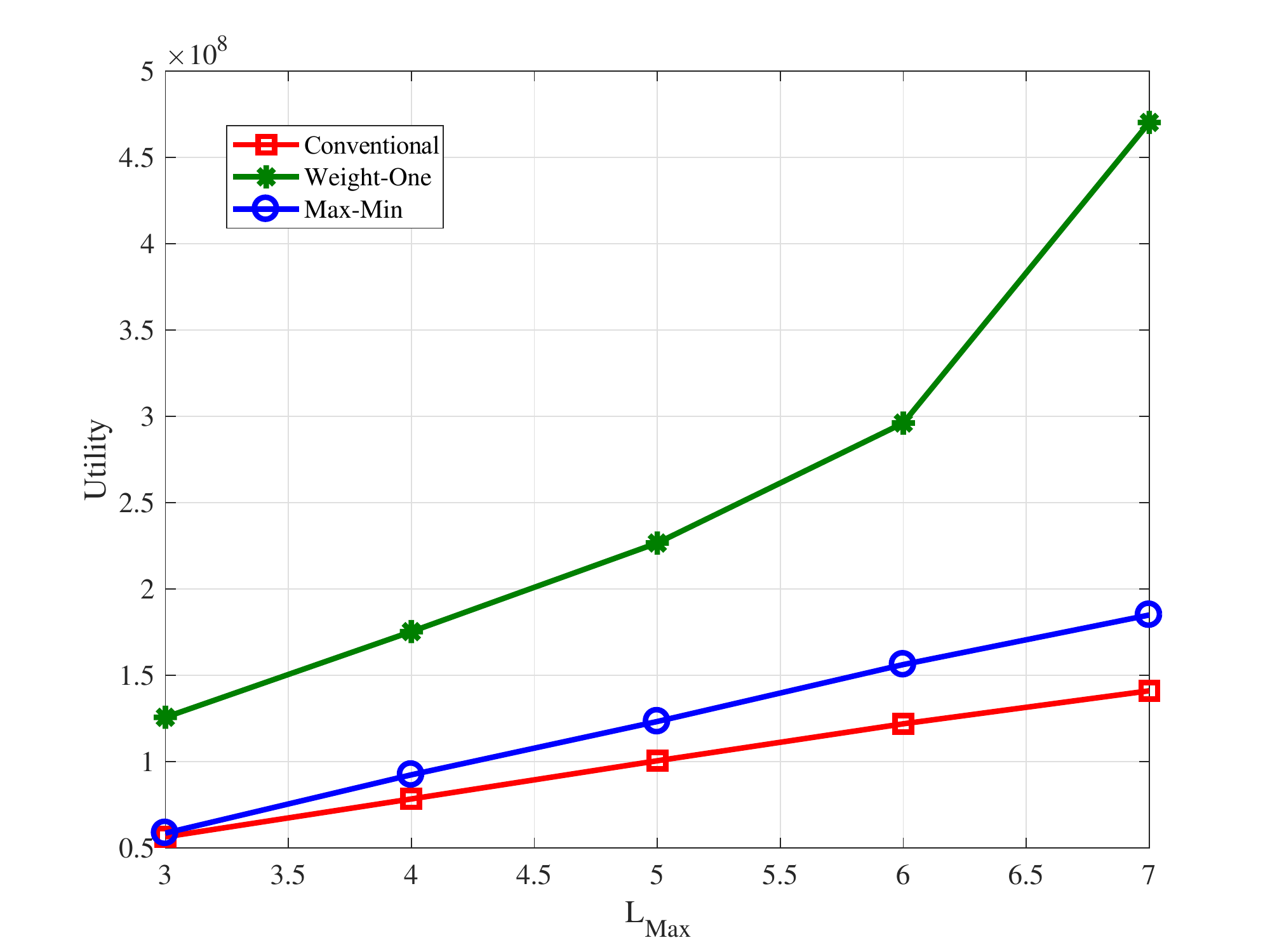}}
	}
	\subfigure{
		{\includegraphics[width=.45\textwidth]{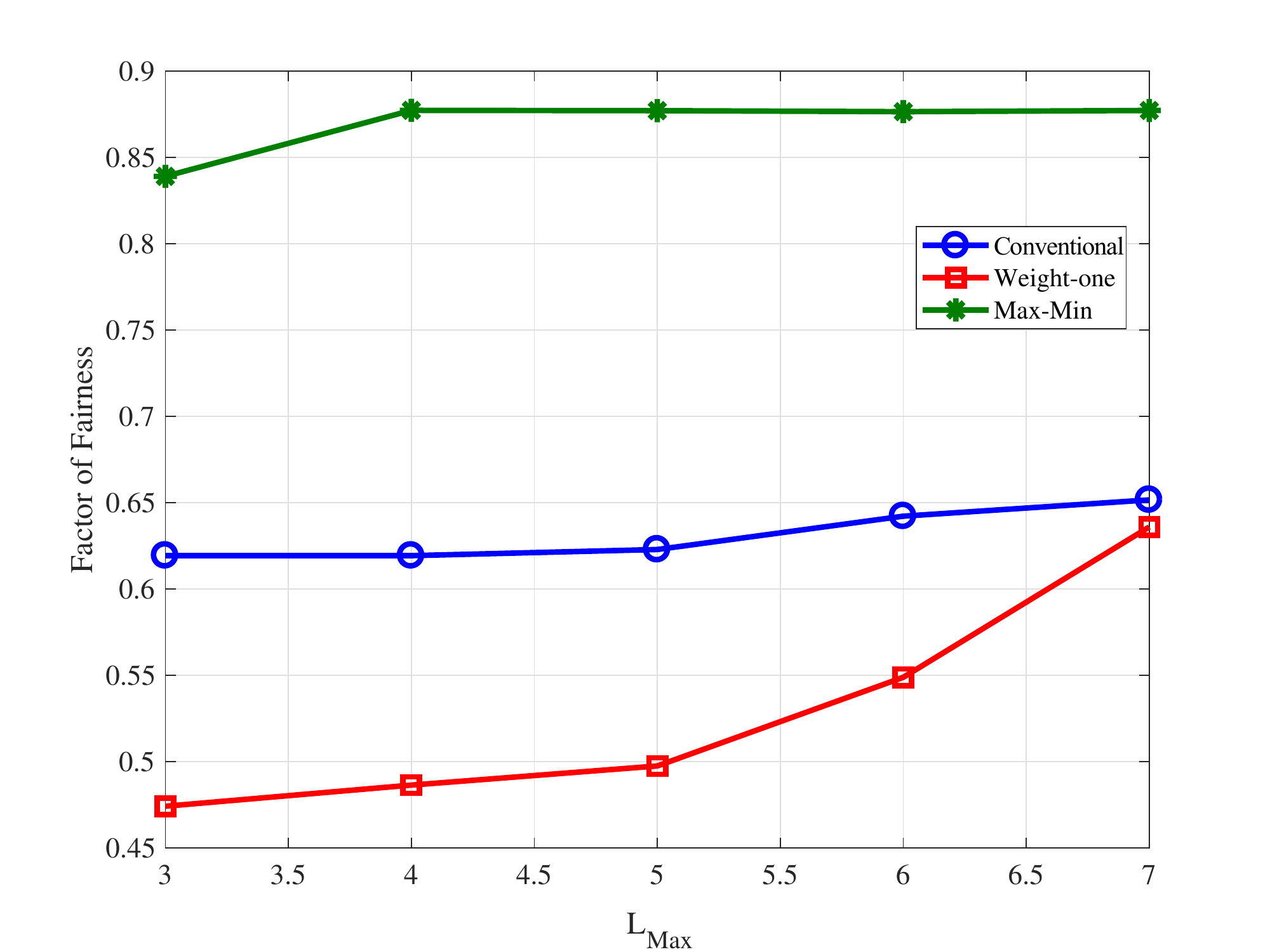}}
	}
	\caption{Left:Summation of players utilities for the proposed methods. Right: Fairness index for the proposed methods.}
	\label{fairness}
\end{figure}

Now using the results in the above figures we obtain Figs.  \ref{fairness}-Right and \ref{fairness}-Left which depict  total revenue and fairness for the max-min, conventional, and weight-one methods.  

In general, we look for a setting which has the best total revenue. However, such revenue has to be divided in a fair fashion among the stakeholders. To quantify this fact, we exploit the Jain fairness index \cite{6547814}. The Jain fairness index for utilities of ISPs, InPs, and SDO  is calculated as follows \cite{6547814}:
\begin{align}
J(\Phi^\text{ISP}_\text{tot},
\Phi^\text{InP}_\text{tot}, 
\Phi^\text{Sens}_\text{tot})=\dfrac{(\Phi^\text{ISP}_\text{tot}+
	\Phi^\text{InP}_\text{tot}+
	\Phi^\text{Sens}_\text{tot})^2}{3((\Phi^\text{ISP}_\text{tot})^2+
	(\Phi^\text{InP}_\text{tot})^2+ 
	(\Phi^\text{Sens}_\text{tot})^2)}.
\end{align}

With the Jain fairness index, when $\Phi^\text{ISP}_\text{tot}=
\Phi^\text{InP}_\text{tot}=
\Phi^\text{Sens}_\text{tot},$ the best fairness (i.e., $J(\Phi^\text{ISP}_\text{tot},
\Phi^\text{InP}_\text{tot}, 
\Phi^\text{Sens}_\text{tot})=1$) is achieved.

As the figures show, if we go with the conventional approach, we are in fact cutting the total revenue. 
Moreover, the degree of fairness is not also acceptable. By using the weight-one approach, the total revenue is drastically increased.  Nevertheless, the resulting fairness is still far from being acceptable. Finally, if we use the max-min approach, we get close to the complete fairness. This is while the total revenue is also increased compared to the conventional approach.

The results prove how the proposed end to end joint pricing and radio resource allocation approach can provide fairness and at the same time, increase the revenue.  This is achieved through jointly optimizing  the parameters corresponding to all players.    

%


%

%

\section{Conclusions}\label{conclusions}

In this paper, we provided a novel comprehensive pricing model for IoT services in 5G networks. We considered all the parties involved in the communication scenario and determined the revenue and cost of each party. 
We formulated the resource allocation in the considered model as a multi-objective optimization problem where in addition to the resource allocation, the pricing variables were also the optimization variables. We solved the resulting problem using the  alternating approach and evaluated the performance of the proposed model for different network scenarios using simulations. Moreover, we presented the conventional approach for pricing and radio resource allocation.
Simulation results indicate that by applying the proposed joint framework, we can increase the total revenue compared to the conventional case while providing an almost complete fairness among the players. This paves the way to reach the stated goal of the paper: removing one of the barriers that prevent IoT from becoming pervasive. 

\begin{figure}
	\centering
	\subfigure{
		{\includegraphics[width=.47\textwidth]{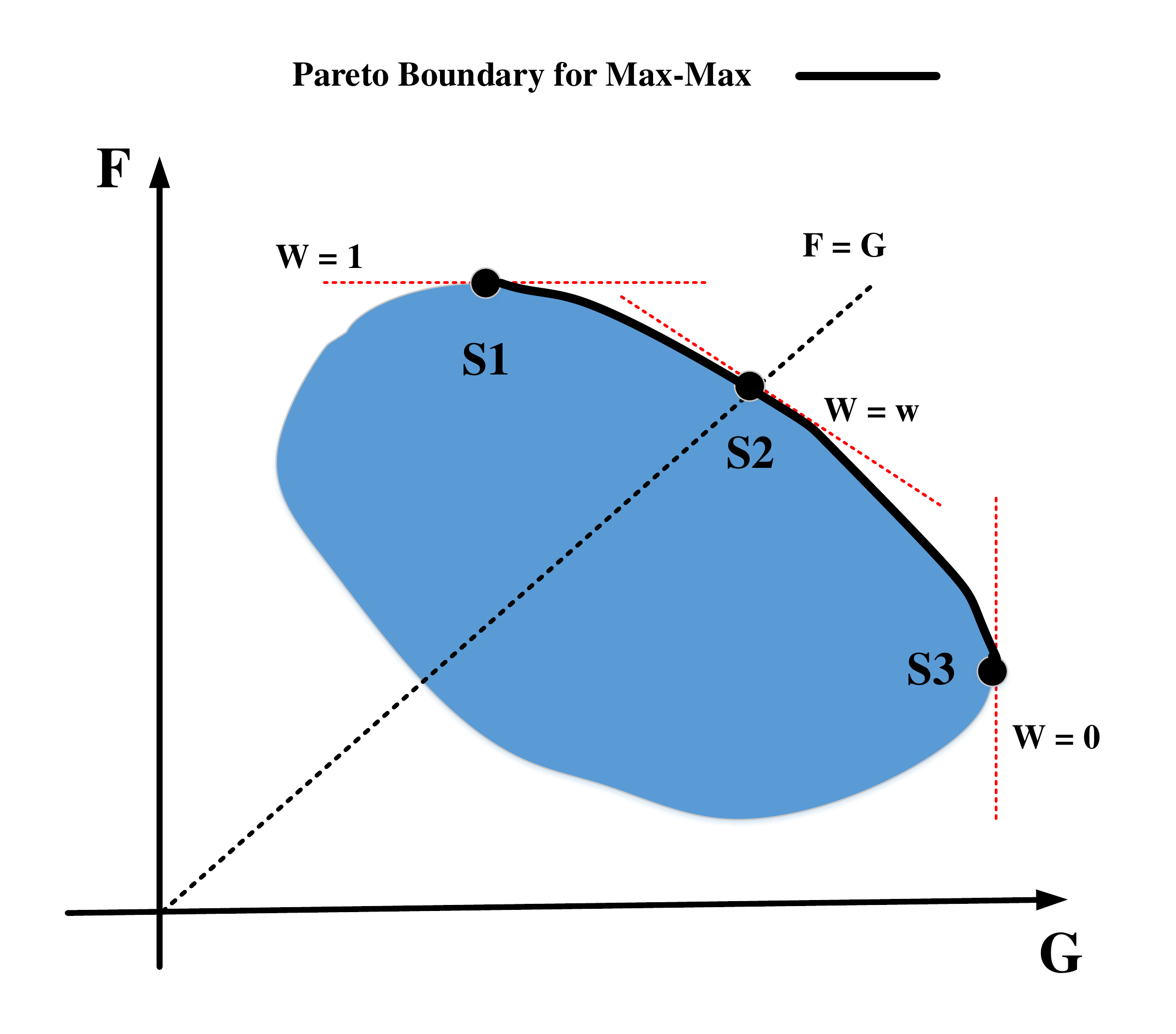}}
	}
	\subfigure{
		{\includegraphics[width=.44\textwidth]{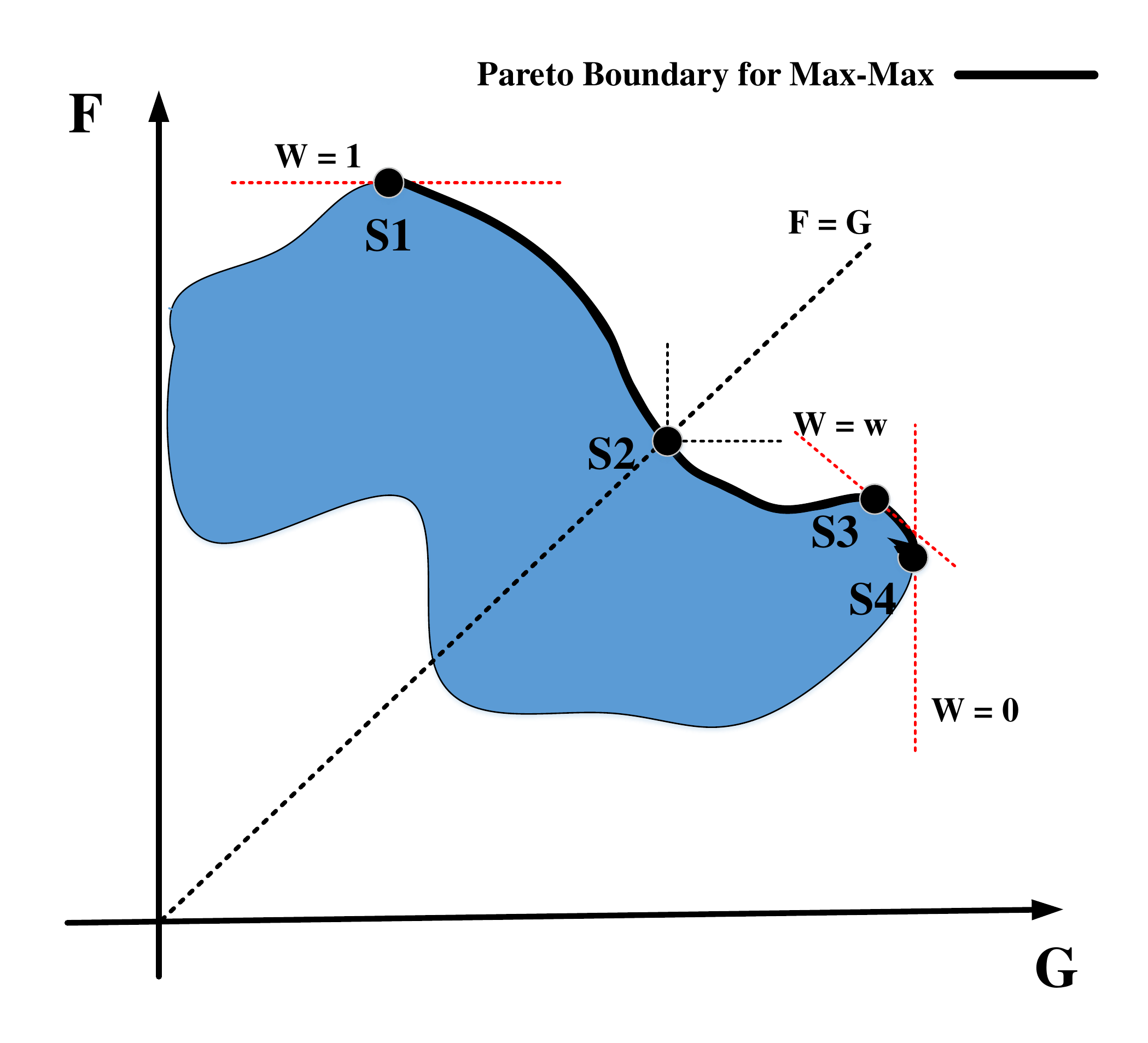}}
	}
	\caption{Left:Convex Pareto boundary of a multi-objective maximization problem with two objective functions. Right: Non-convex Pareto boundary of a multi-objective maximization  problem with two objective functions.}
	\label{Boundry1}
\end{figure}

\appendix
\textbf{Appendix A:} Suppose that we have a multi-objective maximization problem with objective functions $F$ and $G$, and variable vector $\boldsymbol{\varrho}$ formulated  as follows:
	\begin{align}\label{ap}
	\max_{ \boldsymbol{\varrho}}
	\{F(\boldsymbol{\varrho}), G(\boldsymbol{\varrho})\},
	\text{s.t.}:\hspace{.25cm}
	C_1\text{-}C_K,\nonumber
	\end{align}
where $C_1,\dots, C_K$ are the constraints of the optimization problem.
 In order to solve \eqref{ap}, one method is scalarization where the original multi-objective problem is transformed into a single objective problem. There are various methods which transform the original problem into a single objective optimization problem. Each single objective optimization problem gives a Pareto optimal solution of the original multi-objective problem. Each of the solutions is determined by  a point in the Pareto solution set of the original problem. For example,  Fig. \ref{Boundry1} shows the Pareto solution set of the maximization problem in which the bold line determines the Pareto boundary.
 Each of the points can be interpreted as the solution of a specific transformed single objective optimization problem. Suppose that the weighted method is used to transfer the multi-objective into a single objective problem. With the weighted method, the optimization problem \eqref{ap} is reformulated as
  	\begin{align}
  	\max_{ \boldsymbol{\varrho}} WF(\boldsymbol{\varrho})+ (1-W)G(\boldsymbol{\varrho}),
  	\text{s.t.}:\hspace{.25cm}
  	C_1\text{-}C_K.
  	\end{align}
The weighted method is an approach of the scalarization method in which with different values of $W$, various single objective optimization problem can be achieved. When the Pareto solution set is a convex set, such as Fig. \ref{Boundry1}-Left, with the weighted method all of the points in the Pareto boundary can be achieved. For example, point $\textbf{S1}$ can be interpreted as a weighted single objective problem in which the weight of function $F$ is $W=1$.  Point $\textbf{S2}$ can be interpreted as a weighted single objective optimization problem 
in which the main goal is maximizing the fairness between functions $F$ and $G$.  Point $\textbf{S3}$ can be interpreted as a weighted single objective optimization problem in which the weight of function $G$ is $1$ or ($W=0$). As we mentioned, with the weighted approach, all points of the Pareto boundary of a convex Pareto solution set can be achieved. However, finding a value of $W$ which gives the best fairness between $F$ and $G$ is so complicated, while by exploiting the max-min approach as 
	\begin{align}\label{ap12}
	\max_{ \boldsymbol{\varrho}} \{\min_{F,G}	\{F(\boldsymbol{\varrho}), G(\boldsymbol{\varrho})\}\},
	\text{s.t.}:\hspace{.25cm}
	C_1\text{-}C_K,
	\end{align}
 the best fairness  can be easily achieved. Moreover, if the Pareto solution set  not be  a convex set, such as Fig. \ref{Boundry1}-Right, the weighted approach can not achieve all of the Pareto boundary points. For example, in  Fig. \ref{Boundry1}-Right, point $\textbf{S2}$ (consider that $\textbf{S2}$ gives the best fairness) can not be achieved with the weighted approach. In this case, the max-min method can be applied to achieve the best fairness between $F$ and $G$. From the above explanation, we can find that in a multi-objective optimization problem, considering the main goal of the multi-objective optimization problem, a single objective optimization problem can be formulated from the main multi-objective problem.
It should be noted that, in addition to the main goal of the transformed optimization problem, its solution complexity can have essential role in selecting the form of the transformed optimization problem. 
The proposed max-min approach  comes from this fact that the fairness among ISPs, SDO and InPs has the maximum values, and also,  the utility function of users  has the maximum value. It should be noted that, due to the this fact that the user utility has a negative value, it is not possible to consider it as an entry of the max-min approach.

%

\hyphenation{op-tical net-works semi-conduc-tor}
\bibliographystyle{IEEEtran}
\bibliography{IEEEabrv,biblio2}
\end{document}